\documentclass[draft]{agujournal}

\draftfalse

\usepackage{amsmath,amssymb,latexsym}  
\usepackage{mathtools}
\usepackage{textcomp}
\usepackage[english]{babel}
\usepackage{hyperref}

\journalname{JGR-Planets}

\newcommand{\mitbf}[1]{
  \hbox{\mathversion{bold}$#1$}}

\begin{document}

%
%


\title{Viscoelastic relaxation within the Moon and the phase lead of its Cassini state}

%
%




\authors{Olivier Organowski \affil{1} and Mathieu Dumberry \affil{1}}
 
\affiliation{1}{Department of Physics, University of Alberta, Edmonton, Alberta, Canada.}






\correspondingauthor{Mathieu Dumberry}{dumberry@ualberta.ca}




\begin{keypoints}
\item Viscoelastic deformation in the lowermost mantle and inner core of the Moon dissipates a part of the lunar rotation energy.
\item We focus on the role of the inner core and show that maximal lead angles of the Cassini state result for viscosities of $10^{13}-10^{14}$ Pa s.
\item The largest dissipation occurs for large inner cores and when the free inner core nutation frequency is close to the precession frequency.
\end{keypoints}

%
%


\begin{abstract}
Analyses of Lunar Laser Ranging data show that the spin-symmetry axis of the Moon is ahead of its expected Cassini state by an angle of $\phi_p$ = 0.27 arcsec.  This indicates the presence of one or more dissipation mechanisms acting on the lunar rotation.  A combination of solid-body tides and viscous core-mantle coupling have been proposed in previous studies.  Here, we investigate whether viscoelastic deformation within a solid inner core at the centre of the Moon can also account for a part of the observed phase lead angle $\phi_p$.  We build a rotational dynamic model of the Cassini state of the Moon that comprises an inner core, a fluid core and a mantle, and where solid regions are allowed to deform viscoelastically in response to an applied forcing.  We show that the presence of an inner core does not change the global monthly Q of the Moon and hence, that the contribution from solid-body tides to $\phi_p$ is largely unaffected by an inner core.  However, we also show that viscoelastic deformation within the inner core, acting to realign its figure axis with that of the mantle, can contribute significantly to $\phi_p$ through inner core-mantle gravitational coupling.  We show that the contribution to $\phi_p$ is largest when the inner core viscosity is in the range of $10^{13}$ to $10^{14}$ Pa s, when the inner core radius is large and when the free inner core nutation frequency approaches a resonance with the precession frequency of $2\pi/18.6$ yr$^{-1}$. 
\end{abstract}

{\bf Plain language summary:} Analyses of the lunar rotation have revealed that there exists one or more mechanisms that act to dissipate a part of the rotational energy.  Previous mechanisms that have been suggested include friction at the boundary between the fluid core and the mantle, and a delayed response of a deformable 'soft' mantle to tidal forces.  Here, we investigate whether the dissipation of rotational energy could also originate from deformation within a soft solid inner core located at the centre of the Moon.  We show that if the inner core is sufficiently large and if its viscosity is sufficiently low, its contribution to the rotational energy dissipation is not negligible.

\section{Introduction}

In the year 1693 the Franco-Italian astronomer Giovanni Domenico Cassini published a set of three empirical laws describing the rotational motion of the Moon \cite[][]{cassini1693}: 1) The Moon is locked in a 1:1 spin-orbit resonance, such that for every orbit around the Earth the Moon rotates once about its own axis; 2) The Moon's spin-symmetry axis is misaligned by a constant angle relative to the ecliptic normal; 3) the orbit normal and the spin-symmetry axis both precess about the ecliptic normal at the same frequency and the three vectors remain coplanar.  The third law describes a configuration referred to as a Cassini state \cite[][]{colombo66,peale69}.  It is this precessing (quasi) coplanar configuration of the orbit normal, lunar mantle rotation axis, and ecliptic normal that is at the centre of our study. 

\begin{figure}
\begin{center}
    \includegraphics[height=5.5cm]{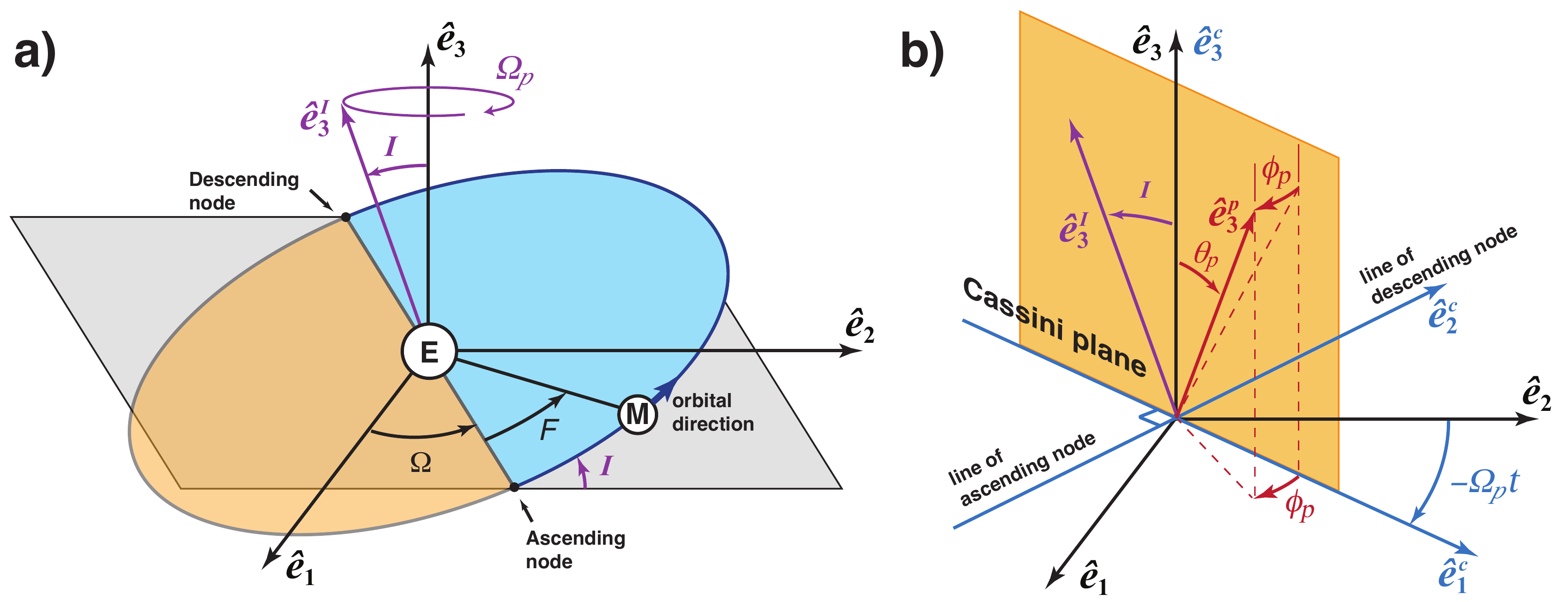} 
    \caption{\label{fig:cassini}  a) The orbit of the Moon (M) around Earth (E) as seen in the inertial frame $(\mitbf{\hat{e}_1}, \mitbf{\hat{e}_2}, \mitbf{\hat{e}_3})$. The normal to the orbital plane is defined by $\mitbf{\hat{e}_3^I}$ and is offset from $\mitbf{\hat{e}_3}$ by an angle $I=5.145^\circ$. $\mitbf{\hat{e}_3^I}$ precesses about $\mitbf{\hat{e}_3}$ in a retrograde direction at frequency $\Omega_p = 2\pi/18.6$ yr$^{-1}$.  $F$ is the mean angle from the orbit's ascending node.  $\Omega$ is the longitude of the ascending node with respect to $\mitbf{\hat{e}_1}$.  The blue (orange) shaded region indicates portions of the orbit when the Moon is above (below) the ecliptic plane, the latter being defined by the vectors $\mitbf{\hat{e}_1}$ and $\mitbf{\hat{e}_2}$ and represented by the grey shade.     b) The Cassini state of the Moon.  The plane (orange shaded area) defined by $\mitbf{\hat{e}_3}$ and $\mitbf{\hat{e}_3^I}$ is the Cassini plane, rotating at frequency $-\Omega_p$ about $\mitbf{\hat{e}_3}$.  The frame attached to this plane is the Cassini frame $(\mitbf{\hat{e}^c_1}, \mitbf{\hat{e}^c_2}, \mitbf{\hat{e}^c_3})$, with $\mitbf{\hat{e}^c_3}$ aligned with $\mitbf{\hat{e}_3}$ and with $\mitbf{\hat{e}^c_2}$ aligned with the line of the descending node.   The symmetry axis of the mantle $\mitbf{\hat{e}_3^p}$ is offset from $\mitbf{\hat{e}_3}$ by $\theta_p=1.543^\circ$.  Without dissipation, $\mitbf{\hat{e}_3^p}$ lies in the Cassini plane.  As a result of dissipation, $\mitbf{\hat{e}_3^p}$ is ahead of the Cassini plane by an angle $\phi_p = 0.27$ arcsec.  a) and b) do not correspond to the same snapshot in time. }
\end{center}
\end{figure}

The orbital geometry and Cassini state configuration of the Moon are illustrated in Figure \ref{fig:cassini}.  The orbit normal  and the symmetry axis precess about the ecliptic normal with a period of 18.6 years.  Detailed observations of the lunar rotation made by Lunar Laser Ranging (LLR) in the decades following the Apollo missions \cite[e.g.][]{dickey94} have revealed that the angle between the orbit and ecliptic normals is $I=5.145^\circ$ and the angle between the symmetry axis and the ecliptic normal is $\theta_p=1.543^\circ$, in the opposite direction of $I$.  In addition, LLR observations indicate that the symmetry axis is not exactly co-planer with the ecliptic and orbit normals, as one would expect for an exact Cassini state, but instead leads ahead of this plane by a small angle of $\phi_p=0.27$ arcsec \cite[][]{yoder81,williams01}. This offset is indicative of rotational energy dissipation within the Moon.

Two dissipation mechanisms have been proposed in the literature. The first is due to solid-body tides \cite[][]{yoder79,cappallo81}.  The Moon deforms in response to changes in the gravitational potential imposed on it primarily by Earth and to a lesser extent by the Sun and other planets.  For a purely elastic deformation, the tidal bulge is in phase with the potential of the tide-raising body on the lunar surface.  However, as a result of solid body dissipation, the phase of the tidal bulge lags behind the potential and this time-lag leads to an associated torque acting on the Moon.  In response to this torque, the spin-symmetry axis of the mantle is displaced in the direction of precession from the plane defined by the ecliptic and orbit normals; a plane which we refer to as the Cassini plane.

The second dissipation mechanism is from viscous friction at the lunar core-mantle boundary (CMB) \cite[][]{yoder81,williams01}. The rotation vector of the fluid core does not follow that of the mantle in its 18.6 yr precession because the ellipticity of the lunar CMB is too small for adequate inertial coupling between the two \cite[][]{goldreich67}. Although the fluid core tilt angle is not known, it is presumed to be close to, though not exactly aligned with, the ecliptic normal \cite[][]{williams01,meyer11,dumberry16,stys18}.  A fluid core that is rotating with a differential velocity relative to the mantle exerts a viscous drag on the CMB, and hence a torque on the mantle which acts to displace its symmetry axis away from the Cassini plane.  

Small physical longitudinal and latitudinal librations of the order of a few milliarcsec provide a pathway to separate the relative contributions from these two dissipation mechanisms.   This is done by comparing the observed changes in the lunar rotation inferred by LLR with the prediction from a rotational model subject to the known torques acting on the Moon \cite[][]{williams01,williams15}.  The rotational Moon model that is used in such LLR studies consists of a rigid solid mantle and a fluid core, but does not include an inner core. Tidal dissipation within the solid mantle is modelled as a time-delay.  Viscous friction at the CMB is modelled with a single dissipation parameter.  The goal of LLR studies is to fit as best as possible the entire time series of the modelled lunar rotation, including all its librations and the equilibrium Cassini state, to the observed data. Evidently, the recovered dissipation parameters are dependent on the assumptions inherent to the rotation model used to fit the LLR observations. If an additional dissipation mechanism exists, its introduction into the rotation model will modify these parameters.

In the present study, we focus on the dissipation associated with the Cassini state which results in the observed phase lead of $\phi_p=0.27$ arcsec.  In particular, we investigate how a third dissipation mechanism may contribute to $\phi_p$: viscoelastic deformation within a solid inner core.  Whether the Moon has a solid inner core remains uncertain.  Thermal evolution models suggest that a solid inner core should have crystallized at its centre \citep{zhang13,laneuville14,scheinberg15}.  A solid inner core with a radius of $240 \pm 10$ km has been inferred based on seismic data \citep{weber11} but this interpretation is not unique \cite[e.g.][]{garcia11}.  If present, the inner core can contribute to the observed lead angle $\phi_p$ in two ways. First, through its solid body viscoelastic deformation, and second, from viscous friction at the inner core boundary (ICB).  We focus on the former.  Viscoelastic deformation within the inner core implies that, as for the mantle, its instantaneous figure axis includes a component out of the Cassini plane.  Because of the inner core's small relative size compared to the mantle, it is unlikely that this can contribute to a significant change in the global tidal deformation of the Moon.  Indeed we confirm this in our study.  However, a more significant effect can arise from the gravitational torque that the mantle and inner core exert on one another's figure.  In the Cassini state, the tilt angle of the inner core with respect to the mantle can be significant \cite[][]{dumberry16,stys18}.  If this tilt includes a component out of the Cassini plane, then the torque it applies on the mantle also includes such an out-of-plane component, thereby contributing to $\phi_p$.

The underlying motivation for our study is three-fold. First, correctly identifying and quantifying the rotational dissipation mechanisms within the Moon is crucial to accurately constrain its rheology.  Indeed, several studies have sought to explain the frequency dependence of the tidal dissipation that is deduced on the basis of LLR observations \cite[e.g.][]{williams01,williams14,williams15,harada14,harada16,khan14,nimmo12,karato13}.  If the inner core participates in the dissipation, the inference from LLR would be changed, and this would then affect the conclusions drawn in these studies.

Second, viscous friction at the CMB between the differentially rotating mantle and fluid core has been suggested as a possible source of energy for the ancient lunar dynamo \cite[][]{williams01,dwyer11,cebron19}.  This inference is based on the fact that the angle of offset between the core and mantle was larger in the past when the Moon was closer to Earth and its tidally locked rotation was also faster.  However, the past dissipation is computed on the basis of the present-day estimate of the viscous friction at the CMB.  If a part of the present-day dissipation includes a contribution from viscous friction at the ICB, or a contribution from a viscously deforming inner core, this will affect the estimate of the power available to drive the ancient lunar dynamo by mechanical forcing.

Third, the tilt angle of the inner core relative to the lunar mantle can theoretically be large \cite[][]{dumberry16,stys18}. In the reference frame of the lunar mantle, a tilted inner core precesses with a period of one lunar day. This precession should manifest itself as a periodic variation in the degree 2 order 1 coefficients of gravity \citep{williams07}.  However, such a gravity signal has not been detected to date \cite[][]{williams15b}.  A possible reason for this is that the viscous relaxation allows the inner core to realign its geometric figure to match the surface of hydrostatic equilibrium imposed by the mantle's gravity field. As was shown in \cite{dumberry16}, if the viscous relaxation timescale of the inner core is of the order of one lunar day, gravitational coupling with the mantle would prevent a misalignment of the  inner core of more than $1^\circ$. Hence, viscous relaxation of the inner core figure may be the reason why its associated gravity signal remains below the detection threshold.


\section{Theory}

\subsection{The interior structure of the Moon}

The model of the Moon's interior structure that we adopt consists of a solid inner core, a fluid outer core, a low seismic velocity transition zone at the base of the mantle, or more succinctly a low velocity zone (LVZ), a solid mantle, and a thin crust. The outer radii of each of these layers, in the same sequence, are denoted as $r_s$, $r_f$, $r_l$, $r_m$, and $R$, and their densities by $\rho_s$, $\rho_f$, $\rho_l$, $\rho_m$, and $\rho_c$. The inner core radius $r_s$ corresponds to the ICB radius, and the fluid core radius $r_f$ corresponds to the CMB radius.  We neglect compressibility effects from increasing pressure with depth (which are small for the Moon) and assume that the density and other material properties within each layer are uniform.  Although crude, adopting uniform layers is broadly consistent with the radial models of the Moon inferred from seismic observations \cite[][]{garcia11,weber11,matsumoto15}.

Each layer is triaxial in shape.   We denote the polar flattening by the variable $\epsilon$, defined as the difference between the mean equatorial and polar radii, divided by the mean spherical radius.  Likewise, we denote the equatorial flattening by the variable $\xi$, defined as the difference between the maximum and minimum equatorial radii, divided by the mean spherical radius.  The set of polar and equatorial flattenings at the outer radius of the inner core, fluid core, LVZ, mantle and at the lunar surface are denoted by ($\epsilon_s$, $\xi_s$), ($\epsilon_f$, $\xi_f$), ($\epsilon_l$, $\xi_l$), ($\epsilon_m$, $\xi_m$) and ($\epsilon_r$, $\xi_r$), respectively.
   
\subsection{The rotational model}

To model the rotational dynamics of the Moon we use the framework developed by \cite{mathews91a} to study the Earth's nutations \cite[see also][]{mathews02,dehant15}.  This framework was adapted to model the Cassini state of the Moon in \cite{dumberry16}, henceforth referred to as DW16.   A further extension of the framework is presented in \cite{stys18},  henceforth referred to as SD18.  We give an outline of this rotational model below, but the interested reader is referred to DW16 and SD18 for more details.  Here, we focus on the new additions to this model in order to include the dissipation involved in the Cassini state.  

From the perspective of the rotational dynamics, the LVZ, mantle and crust are welded together and form a single rotating region.  In the context of the rotating model we refer to this region as the ``mantle'', but it should not obscure the fact that material properties may be distinct in each of the crust, LVZ and the mantle in between.

\begin{figure}
\begin{center}
    \includegraphics[height=7cm]{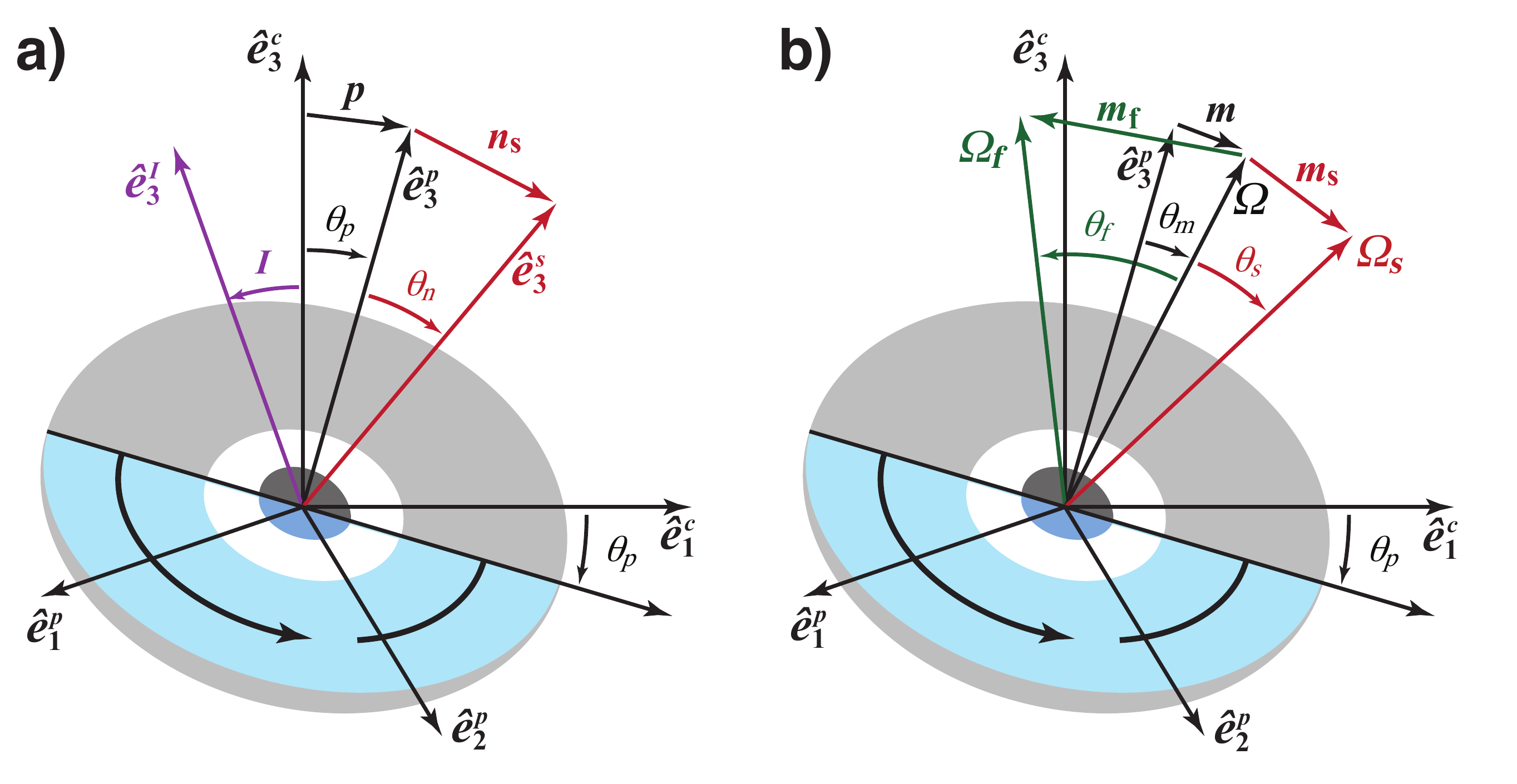} 
    \caption{\label{fig:angles}  The orientation of (a) the orbit normal ($\mitbf{\hat{e}_3^I}$), the symmetry axes of the mantle ($\mitbf{\hat{e}^p_3}$) and inner core ($\mitbf{\hat{e}^s_3}$), and (b) the rotation rate vectors of the mantle ($\mitbf{\Omega}$), fluid core ($\mitbf{\Omega_f}$) and inner core ($\mitbf{\Omega_f}$) as seen in the Cassini frame ($\mitbf{\hat{e}^c_1}$, $\mitbf{\hat{e}^c_3}$).  Also shown are the vectors ${\bf p}$, ${\bf n_s}$, ${\bf m}$, ${\bf m_f}$ and ${\bf m_s}$ and associated angles $\theta_p$, $\theta_n$, $\theta_m$, $\theta_f$ and $\theta_s$.  For ease of illustration, all vectors are shown to lie in the Cassini plane (defined by $\mitbf{\hat{e}^c_3}$ and $\mitbf{\hat{e}^I_3}$), corresponding to a situation with no dissipation.  When dissipation is present, all vectors have a component out of this plane.   The light grey, white, and dark grey ellipsoid represent a polar cross-section of the mantle, fluid core and inner core, respectively.  Blue shaded parts show the equatorial cross section.  The black curved arrow in the equatorial plane indicates the direction of rotation, at frequency $\Omega_o+\Omega_p \cos(\theta_p) = 2 \pi / 27.212$ day$^{-1}$ of the mantle axes $\mitbf{\hat{e}^p_1}$ and $\mitbf{\hat{e}^p_2}$ about the Cassini plane. }
    \end{center}
\end{figure}

As shown in Figure \ref{fig:cassini}, the Cassini state of the Moon can be defined in terms of two reference frames: the inertial frame, defined by unit vectors ($\mitbf{\hat{e}_1}$, $\mitbf{\hat{e}_2}$, $\mitbf{\hat{e}_3}$), with $\mitbf{\hat{e}_3}$ aligned with the ecliptic normal; and the Cassini frame specified by unit vectors $(\mitbf{\hat{e}^c_1}, \mitbf{\hat{e}^c_2}, \mitbf{\hat{e}^c_3})$, with $\mitbf{\hat{e}^c_3}$ aligned with $\mitbf{\hat{e}_3}$.  The Cassini frame is defined in terms of the orientation of the orbit normal $\mitbf{\hat{e}^I_3}$ with respect to the ecliptic normal $\mitbf{\hat{e}_3}$:  $\mitbf{\hat{e}^c_3}$ is set to be aligned with $\mitbf{\hat{e}_3}$, and the orientation of $\mitbf{\hat{e}^c_1}$ is set to lie in the plane containing  $\mitbf{\hat{e}^I_3}$ and $\mitbf{\hat{e}_3}$, which we refer to as the Cassini plane.  The Cassini frame is rotating with frequency $-\Omega_p$ about $\mitbf{\hat{e}_3}$ with respect to the inertial frame.  

The rotational model that we use is defined with respect to a third reference frame attached to the rotating mantle, with unit vectors $(\mitbf{\hat{e}^p_1}, \mitbf{\hat{e}^p_2}, \mitbf{\hat{e}^p_3})$.   $\mitbf{\hat{e}^p_3}$ is chosen to be aligned with the maximum (polar) moment of inertia of the mantle and consequently it defines the orientation of the symmetry (or figure) axis of the mantle.  $\mitbf{\hat{e}^p_1}$ and $\mitbf{\hat{e}^p_2}$ are aligned, respectively, with the minimum and intermediate moments of inertia (both in equatorial directions).  Since the Moon is in a tidally locked 1:1 spin-orbit resonance, $\mitbf{\hat{e}^p_1}$ is at the central point of the hemisphere of the Moon facing the Earth (near side) and, when averaged over one full orbit, points directly towards Earth.  The reader is referred to Appendix A of SD18 for a more in-depth discussion of the links between the inertial ($\mitbf{\hat{e}}$), Cassini ($\mitbf{\hat{e}^c}$) and mantle ($\mitbf{\hat{e}^p}$) frames.  The misalignment of the lunar mantle's symmetry axis $\mitbf{\hat{e}^p_3}$ relative to the ecliptic normal $\mitbf{\hat{e}_3}$ is described by a vector ${\bf p}$, 

\begin{subequations}
\begin{equation}
\mitbf{\hat{e}^p_3} = \mitbf{\hat{e}_3}  + {\bf p} \, .
\end{equation}

A similar coordinate system is defined for the inner core.  Let ($\mitbf{\hat{e}_1^s}$, $\mitbf{\hat{e}_2^s}$, $\mitbf{\hat{e}_3^s}$) be the unit vectors in the direction of the inner core's principle moments of inertia, defined in the same way as for the lunar mantle. Because the inner core is suspended within the fluid core, it is free to take on a different orientation than the mantle.  The misalignment of the inner core symmetry axis $\mitbf{\hat{e}_3^s}$ relative to the mantle symmetry axis $\mitbf{\hat{e}_3^p}$ is described by a vector ${\bf n_s}$, such that

\begin{equation}
\mitbf{\hat{e}^s_3} = \mitbf{\hat{e}^p_3} + {\bf n_s} = \mitbf{\hat{e}_3}  + {\bf p} + {\bf n_s}  \, .
\end{equation}
\end{subequations}

The rotation and symmetry axes of the mantle -- and similarly those of the inner core -- are expected to remain in close alignment, but they do not coincide exactly.  The misalignment of the mantle rotation rate vector $\boldsymbol{\Omega}$ relative to the mantle symmetry axis $\mitbf{\hat{e}^p_3}$ is captured by a vector ${\bf m}$, such that

\begin{subequations}
\begin{equation}
\boldsymbol{\Omega} = \boldsymbol{\Omega_o} + \boldsymbol{\omega_m} = \Omega_o ( \mitbf{\hat{e}^p_3} + {\bf m} ) \, ,
\end{equation}
where $\Omega_o = 2\pi/27.322$ day$^{-1}$ is the sidereal frequency of the rotation of the lunar mantle. The vector $\boldsymbol{\omega_m} =  \Omega_o {\bf m}$ is the differential angular velocity of the mantle defined with respect to $\mitbf{\hat{e}^p_3}$. The rotation rate vectors of the inner core, $\boldsymbol{\Omega_s}$, and the fluid core, $\boldsymbol{\Omega_f}$ are defined in terms of vectors ${\bf m_s}$ and ${\bf m_f}$, 

\begin{align}
\boldsymbol{\Omega_s} & = \boldsymbol{\Omega} + \boldsymbol{\omega_s}  = \Omega_o (  \mitbf{\hat{e}^p_3}+ {\bf m} + {\bf m_s} ) \, ,\\
\boldsymbol{\Omega_f} & = \boldsymbol{\Omega} + \boldsymbol{\omega_f} = \Omega_o 
( \mitbf{\hat{e}^p_3}+ {\bf m} + {\bf m_f} ) \, ,
\end{align}
\end{subequations}
where $\boldsymbol{\omega_s} =  \Omega_o {\bf m_s}$ and $\boldsymbol{\omega_f} =  \Omega_o {\bf m_f}$ are the differential angular velocities of the inner core and fluid core, respectively, both defined with respect to the mantle rotation rate vector. The definitions of $\boldsymbol{\Omega}$, $\boldsymbol{\Omega_s}$ and $\boldsymbol{\Omega_f}$ are sufficiently accurate for our purpose here, but more proper definitions are given in Appendix A of SD18.

A graphical depiction of the five rotational vectors ${\bf p}$, ${\bf m}$, ${\bf m_f}$, ${\bf m_s}$ and ${\bf n_s}$ is shown in Figure \ref{fig:angles}.  They are the five unknowns of our system which are found by solving a system of five coupled equations.  The first three equations describe, respectively,  the time rate of change of the angular momenta of the whole Moon (${\bf H}$), the fluid core (${\bf H_f}$), and the inner core (${\bf H_s}$) in the reference frame of the rotating mantle (DW16),

\begin{subequations}
\begin{align}
 \frac{d}{dt} {\bf H} + \boldsymbol{\Omega} \times {\bf H} & = \boldsymbol{\Gamma_e}  +\boldsymbol{\Gamma_t} \, , \label{eq:dHdt} \\
\frac{d}{dt} {\bf H_f} - \boldsymbol{\omega_f} \times {\bf H_f} &= - \boldsymbol{\Gamma_{cmb}} - \boldsymbol{\Gamma_{icb}} \, ,\label{eq:dHfdt} \\
\frac{d}{dt} {\bf H_s} + \boldsymbol{\Omega} \times {\bf H_s} & =  \boldsymbol{\Gamma_{es}}+\boldsymbol{\Gamma_{ts}}+ \boldsymbol{\Gamma_s} + \boldsymbol{\Gamma_{icb}} \, .\label{eq:dHsdt}
\end{align}
Here, $\boldsymbol{\Gamma_e}$ and $\boldsymbol{\Gamma_{es}}$ are the gravitational torques from Earth acting on the whole Moon and on the inner core, respectively.  $\boldsymbol{\Gamma_{t}}$ and $\boldsymbol{\Gamma_{ts}}$ are the torques associated with the tidal dissipation of the whole Moon and inner core, respectively. $\boldsymbol{\Gamma_s}$ is the torque from pressure and self-gravitation exerted on the inner core, and $\boldsymbol{\Gamma_{icb}}$ and $\boldsymbol{\Gamma_{cmb}}$ are the torques from surface tractions on the inner core (at the ICB) and on the fluid core (at the CMB), respectively.  The remaining two equations are kinematic relationships, one to express the change in the orientation of the inner core figure resulting from its own rotation, and a second describing the invariance of the ecliptic normal in the inertial frame as seen in the mantle frame (SD18).  They are, respectively, 

\begin{align}
& \frac{d}{dt}  \mitbf{\hat{e}^s_3}+\mitbf{\hat{e}^s_3}  \times \boldsymbol{\omega_s}= {\bf 0}\, ,\label{eq:de3sdt}\\
& \frac{d}{dt}  \mitbf{\hat{e}_3}  + \boldsymbol{\Omega} \times  \mitbf{\hat{e}_3}  = {\bf 0} \, .\label{eq:de3dt}
\end{align}
\label{eq:sys}
\end{subequations}

The three angular momentum vectors are expanded in terms of the moment of inertia tensors for the whole Moon ($\boldsymbol{\cal I}$), fluid core ($\boldsymbol{{\cal I}_f}$), and inner core ($\boldsymbol{{\cal I}_s}$) as follows

\begin{subequations}
\begin{align}
    {\bf H} &= \boldsymbol{\cal I} \cdot \boldsymbol{\Omega} + \boldsymbol{{\cal I}_f} \cdot \boldsymbol{\omega_f} + \boldsymbol{{\cal I}_s} \cdot \boldsymbol{\omega_s} \, , \\
     {\bf H_f} & = \boldsymbol{{\cal I}_f} \cdot \boldsymbol{\Omega_f} \, , \\
      {\bf H_s} & = \boldsymbol{{\cal I}_s} \cdot \boldsymbol{\Omega_s} \, .
\end{align}
\end{subequations} 
The moments of inertia tensors involve the principal moments of inertia of the whole Moon ($C>B>A$), fluid core ($C_f>B_f>A_f$) and solid inner core ($C_s>B_s>A_s$).  $C$, $C_f$ and $C_s$ are the polar moments of inertia of each region.   The mean equatorial moments of inertia of each region are defined as

\begin{equation}
\bar{A} = \frac{1}{2} (A+B)  \, ,\hspace*{0.5cm} \bar{A}_f = \frac{1}{2} (A_f+B_f)  \, ,\hspace*{0.5cm} \bar{A}_s = \frac{1}{2} (A_s+B_s)   \, .
\end{equation}
The dynamical ellipticities for the whole Moon ($e$), the fluid core ($e_f$), and the inner core ($e_s$), are then defined by 

\begin{equation}
e = \frac{C - \bar{A}}{\bar{A}} \, , \hspace*{0.5cm} e_f = \frac{C_f - \bar{A}_f}{\bar{A}_f} \, , \hspace*{0.5cm} e_s = \frac{C_s - \bar{A}_s}{\bar{A}_s} \, .
\end{equation}
  
The moment of inertia tensors are defined as

\begin{subequations}
\begin{align}
\boldsymbol{\cal I} & = \bar{A}\boldsymbol{\mathsf{I}} + \bar{A} e \, \mitbf{\hat{e}^p_3} \mitbf{\hat{e}^p_3} +  \alpha_3 \bar{A}_s e_s  (\mitbf{\hat{e}^s_3} \mitbf{\hat{e}^s_3} - \mitbf{\hat{e}^p_3} \mitbf{\hat{e}^p_3}) + \sum_{ij} c_{ij} \mitbf{\hat{e}^p_i} \mitbf{\hat{e}^p_j} \, , \label{eq:Itensormoon} \\
\boldsymbol{{\cal I}_f} & = \bar{A}_f \boldsymbol{\mathsf{I}} + \bar{A}_f e_f \, \mitbf{\hat{e}^p_3} \mitbf{\hat{e}^p_3} +  \alpha_1 \bar{A}_s e_s  (\mitbf{\hat{e}^p_3} \mitbf{\hat{e}^p_3} - \mitbf{\hat{e}^s_3} \mitbf{\hat{e}^s_3}) + \sum_{i,j} c^f_{ij} \mitbf{\hat{e}^p_i} \mitbf{\hat{e}^p_j} \, , \\
\boldsymbol{{\cal I}_s} & = \bar{A}_s \boldsymbol{\mathsf{I}}  +  \bar{A}_s e_s \,
\mitbf{\hat{e}^s_3} \mitbf{\hat{e}^s_3} + \sum_{i,j} c^s_{ij} \mitbf{\hat{e}^p_i} \mitbf{\hat{e}^p_j} \, ,
\end{align}
\label{eq:Itensor}
\end{subequations}
where, for uniform density layers,

\begin{equation}
 \alpha_1=\frac{\rho_f}{\rho_s} \, , \hspace*{1cm} \alpha_3 =1-\alpha_1 =1 - \frac{\rho_f}{\rho_s} \, ,
 \end{equation}
and where $\boldsymbol{\mathsf{I}} $ is the unit tensor. In the reference frame of the lunar mantle, $\boldsymbol{\mathsf{I}}  = \mitbf{\hat{e}^p_i} \mitbf{\hat{e}^p_j} \delta_{ij}$, where $\delta_{ij}$ is the Kronecker delta. The quantities $c_{ij}$, $c^f_{ij}$ and $c^s_{ij}$ represent perturbations to the moment of inertia tensors resulting from viscoelastic deformations \cite[][]{mathews91a,buffett93}.

Viewed in the $\mitbf{\hat{e}^p}$-frame attached to the rotating mantle, the Cassini plane is rotating in a retrograde direction about $\mitbf{\hat{e}_3}$ (see Figure \ref{fig:angles}).  The frequency of its rotation is $\omega \Omega_o$, where $\omega$, expressed in cycles per lunar day, is equal to 

\begin{subequations}
\begin{equation}
\omega = -1 - \delta \omega \cos (\theta_p) \, ,
\label{eq:omega}
\end{equation}
as specified in SD18, although here we follow the approximation ($\cos(\theta_p) \rightarrow 1$) used in DW16 and take

\begin{equation}
\omega = -1 - \delta \omega \, .
\label{eq:omega}
\end{equation}
\end{subequations}
The factor $\delta \omega = \Omega_p/\Omega_o $= 27.322 days / 18.6 yr = $4.022 \times 10^{-3}$ is the Poincar\'e number, expressing the ratio of lunar precession to lunar rotation frequencies.  In the absence of dissipation, the gravitational torque by the Earth on the Moon ($\boldsymbol{\Gamma_e}$) is the only external torque driving a rotational response.  Averaged over one orbit, this torque is directed along the line of the ascending node of the lunar orbit, perpendicular to the Cassini plane in direction $- \mitbf{\hat{e}^c_2}$.  Hence, viewed from the $\mitbf{\hat{e}^p}$-frame, the orientation of this average torque is also rotating at frequency $\omega \Omega_o$.  Setting the equatorial directions $\mitbf{\hat{e}_1^p}$ and $\mitbf{\hat{e}_2^p}$ to correspond to the real and imaginary axes of the complex plane, respectively, we can write the equatorial components of this periodic applied torque in a compact form as 

\begin{subequations}
\begin{equation}
{\Gamma}_1(t) + i {\Gamma}_2(t) = - i \Omega_o^2 \bar{A} \,  \tilde{\phi} \,  \exp[{i \omega \Omega_o t}] \, , \label{eq:gammaphi}
\end{equation}
where 

\begin{equation}
\tilde{\phi} \equiv \tilde{\phi}(\omega \Omega_o) = \phi_{1}(\omega \Omega_o) + i \, \phi_{2}(\omega \Omega_o) \, , 
\end{equation}
\end{subequations}
represents the complex non-dimensional amplitude of the torque rotating at frequency $\omega \Omega_o$, with  $\phi_{1}$ and $\phi_{2}$ its real and imaginary components, respectively.

The five rotational vectors ${\bf p}$, ${\bf m}$, ${\bf m_f}$, ${\bf m_s}$ and ${\bf n_s}$ capture the response of the Moon to this applied torque.  In the absence of dissipation, each of these vectors lie in the Cassini plane but they have different orientations.  Viewed in the $\mitbf{\hat{e}^p}$-frame, their time-dependency is then also proportional to $\exp[{i \omega \Omega_o t}]$, and their two equatorial components can likewise be written as complex variables in the complex plane, in terms of a product between a complex amplitude  and a periodic time-dependency $\exp[{i \omega \Omega_o t}]$.  For instance, ${\bf m}$ is written as 

\begin{subequations}
\begin{equation}
m_1(t) + i m_2(t) =  \tilde{m} \,  \exp[{i \omega \Omega_o t}] \, , 
\end{equation} 
where $\tilde{m}$ is the complex amplitude at frequency $\omega \Omega_o$,

\begin{equation}
\tilde{m} \equiv \tilde{m}(\omega \Omega_o) = m_1(\omega \Omega_o) + i \, m_2(\omega \Omega_o) \, .  
\end{equation} 
\end{subequations}
The other four rotational variables are similarly written in terms of their complex amplitudes $\tilde{p}$, $\tilde{m}_f$, $\tilde{m}_s$ and $\tilde{n}_s$.  Small amplitudes are assumed, in which case $\tilde{p}$, $\tilde{m}$, $\tilde{m}_f$, $\tilde{m}_s$ and $\tilde{n}_s$ are equivalent to the angles of misalignment $\theta_p$, $\theta_m$, $\theta_f$, $\theta_s$ and $\theta_n$, respectively, as shown in Figure \ref{fig:angles}.  

Under the small angle assumption, only the components $c_{13}\mitbf{\hat{e}^p_1}$ and $c_{23} \mitbf{\hat{e}^p_2}$ of the moment of inertia tensor of the whole Moon in Equation (\ref{eq:Itensormoon}) are retained \cite[][]{mathews91a}.  Because the applied torque is periodic, the perturbation in the moment of inertia is also periodic.  This time dependent deformation is cast into our complex notation as

\begin{subequations}
\begin{equation}
c_{13}(t) + i c_{23}(t) = \tilde{c} \, \exp[{i \omega \Omega_o t}] \, ,\label{eq:c123t}
\end{equation} 
where 

\begin{equation}
\tilde{c}\equiv \tilde{c}(\omega \Omega_o) = c_{13}(\omega \Omega_o) + i c_{23} (\omega \Omega_o)
\end{equation} 
\end{subequations}
is the complex amplitude of the moment of inertia perturbation at frequency $\omega \Omega_o$.  Equivalent definitions apply to the perturbations in the moments of inertia of the fluid core and inner core, with $\tilde{c}_f$ and $\tilde{c}_s$ denoting their complex amplitudes, respectively.  (Note that our notation is different from the one used in \cite{mathews91a}, where these were denoted instead by $\tilde{c}_3$, $\tilde{c}_3^f$ and $\tilde{c}_3^s$.) 

We must choose a reference time $t=0$ to orient the Cassini frame with the inertial frame.  We set $t=0$ to correspond to when the line of the ascending node coincides with $\mitbf{\hat{e}_1}$ in Figure \ref{fig:cassini} (i.e. when the longitude of the ascending node is $\Omega=0$, and hence when $\mitbf{\hat{e}^c_1}$ is aligned with $\mitbf{\hat{e}_2}$ (see Figure  \ref{fig:cassini}b)).  We must also choose a mean angle $F$ in Figure \ref{fig:cassini}a at the reference time $t=0$, which is equivalent to choosing the orientation between the mantle frame with the Cassini frame.  We pick $t=0$ to correspond to $F=3\pi/2$ or, equivalently, $F=-\pi/2$.  At this point in the lunar orbit, $\mitbf{\hat{e}^p_1}$ is aligned with $\mitbf{\hat{e}^c_1}$, and the applied torque as seen in the mantle frame is directed towards $-\mitbf{\hat{e}^p_2}$, or in the negative imaginary direction of the complex plane.  With this choice, the complex amplitude of the torque $\tilde{\phi}$ defined in Equation (\ref{eq:gammaphi}) is purely real and positive.  In the absence of dissipation, the response of the Moon is perfectly in phase with the applied torque: the complex amplitudes of all rotational variables ($\tilde{p}$, $\tilde{m}$, etc.) are then also purely real and they all lie on the Cassini plane.   When dissipation is present, at the same point in the lunar orbit, the different rotational vectors no longer lie exactly on the Cassini plane.  That is, their complex amplitudes have both a real and an imaginary component.  The real part represents the response that is in-phase with the applied torque, and the imaginary part the out-of-phase response.  The latter is indicative of dissipation.

The half-period modulation of the gravitational torque by Earth over one orbit and the eccentricity of the orbit lead to small latitudinal and longitudinal librations of the Moon in space.  Likewise, modulations of the gravitational torque by the Sun and other planets also induce librations.  These are neglected in our study, as we focus on the Cassini state equilibrium.   

\subsection{Moment of inertia tensor perturbations}

The perturbations in the moment of inertia tensors $\tilde{c}$, $\tilde{c}_f$, and $\tilde{c}_s$ can be split into two components: an external component from the gravitational potential imposed by Earth (denoted with a superscript $e$); and an internal component (superscript $i$) from the changes in the centrifugal potential and interior mass distribution of the Moon,

\begin{equation}
\tilde{c} = \tilde{c}^{\,e} +  \tilde{c}^{\,i} \, , \hspace*{0.5cm} \tilde{c}_f = \tilde{c}_f^{\, e} +  \tilde{c}_f^{\, i}\, , \hspace*{0.5cm} \tilde{c}_s = \tilde{c}_s^{\,e} +  \tilde{c}_s^{\,i} 
\, . \label{eq:tildec}
\end{equation}

Each of these can be expressed as a linear combination of the rotation variables and a set of compliances.  Following the notation introduced by \cite{buffett93}, we denote these compliances by $S_{ij}$.   The perturbations in the moment of inertia tensors from internal contributions are defined as

\begin{subequations}
\begin{align}
\tilde{c}^{\,i} & = \bar{A} \big( S_{11} \tilde{m}  + S_{12} \tilde{m}_f + S_{13} (\tilde{m}_s - \tilde{\phi}_s^{\, i})  \big) \, ,\\
\tilde{c}_f^{\,i}  & = \bar{A}_f \big( S_{21} \tilde{m}  + S_{22} \tilde{m}_f + S_{23} (\tilde{m}_s - \tilde{\phi}_s^{\, i}) \big) \, , \\
\tilde{c}_s^{\,i}  & = \bar{A}_s \big( S_{31} \tilde{m}  + S_{32} \tilde{m}_f + S_{33} (\tilde{m}_s - \tilde{\phi}_s^{\, i}) \big) \, , 
\end{align}
\end{subequations} 
where $\tilde{\phi}_s^{\,i}= \tilde{\phi}_s^{c}+\tilde{\phi}_s^{g}$  is the sum of the centrifugal potential ($\tilde{\phi}_s^{c}$) at the ICB and the gravitational potential from the rest of the Moon ($\tilde{\phi}_s^{g}$) acting on a tilted inner core.  The latter two can be written as 

\begin{equation}
   \tilde{\phi}_s^{c} = - \alpha_1 \tilde{n}_s \, , \hspace*{0.5cm} \tilde{\phi}_s^{g} = \alpha_3 \alpha_g \tilde{n}_s \, , \label{eq:phiscg}
\end{equation}
where the coefficient $\alpha_g$ captures the strength of the gravitational coupling between a tilted inner core and the rest of the Moon \cite[][]{mathews91a}.  For uniform density layers, with no density contrast at the boundary between the mantle and the LVZ, it is given by

\begin{equation}
\alpha_g  = \frac{8\pi G}{5\Omega_o^2} \left[ \rho_c (\epsilon_r - \epsilon_m) + \rho_m (\epsilon_m - \epsilon_f) + \rho_f \epsilon_f \right] \, ,
\end{equation}
where $G$ is the gravitational constant. Defining the parameter $\alpha_2$ as

\begin{equation}
\alpha_2 = \alpha_1 - \alpha_3 \alpha_g  \, ,
\end{equation}
we can write 

\begin{equation}
    \tilde{\phi}_s^{\,i}= - (\alpha_1 - \alpha_3 \alpha_g)  \tilde{n}_s  = - \alpha_2  \tilde{n}_s \, .
\end{equation}

In the absence of dissipation, the time-dependent perturbation in the moment of inertia tensor resulting from the external gravitational potential from Earth (mass $M_E$) is given by \cite[e.g.][Equation 7]{williams01}

\begin{equation}
c^{\,e}_{ij}(t) = - k_2 \frac{M_E R^5}{r^3} \Big( u_i u_j - \frac{\delta_{ij}}{3} \Big) \, , \label{eq:cek2}
\end{equation}
where $k_2$ is the degree 2 tidal Love number, $r$ is the distance to Earth and $u_i$  are the components of the unit vector of Earth's position as seen in the mantle frame.  The torque by the Earth on the Moon associated with the Cassini state is described in the next section, but averaged over one orbit, it is equivalent to that produced by a mass $M_E/2$ located at a distance $r= a_L \sqrt{1-e_L^2}$, where $a_L$and $e_L$ are the semi-major orbital axis and orbit eccentricity of the Moon, and with a periodic time-dependent position given by 

\begin{equation}
{\bf u} = \cos(I+\theta_p) \Big(  \cos(\omega \Omega_o t) \mitbf{\hat{e}_1^p} + \sin(\omega \Omega_o t) \mitbf{\hat{e}_2^p} \Big) + \sin(I + \theta_p)  \mitbf{\hat{e}_3^p} \, . \label{eq:u}
\end{equation}

The periodic tidal potential associated with the Cassini state results from the same equivalent mass.  Replacing $M_E$ by $M_E/2$, setting $r= a_L \sqrt{1-e_L^2}$, and using Equation (\ref{eq:u}) in Equation (\ref{eq:cek2}), the periodic tidal perturbation in the moment of inertia can be written in the form of Equation (\ref{eq:c123t}), with an amplitude  $\tilde{c}^{\,e}$ at frequency $\omega\Omega_o$ given by,

\begin{equation}
\tilde{c}^{\,e} = - \frac{k_2 M_E R^5}{2 a_L^3 (1-e_L^2)^{3/2}}\cos(I+\theta_p) \sin(I+\theta_p)  \, .
\end{equation}
The Love number $k_2$ is connected to the real part of the compliance $S_{11}$ by 

\begin{equation}
k_2 = \frac{3 G  \bar{A}}{R^5 \Omega_o^2} \cdot Re[S_{11}] \,,  \label{eq:k2}
\end{equation}
and under the assumption of a small angle $\theta_p$ and substituting $\theta_p\approx \tilde{p}$, and using Equation (A26a) of DW16, we can express  $\tilde{c}^{\,e}$ as

\begin{equation}
\tilde{c}^{\,e} = - \bar{A} \cdot  Re[S_{11}] \Big( \Phi_1 + \Phi_2 \tilde{p} \Big) \, , \label{eq:ce1}
\end{equation}
where the  factors $\Phi_1$ and $\Phi_2$ are given by 

\begin{subequations}
\begin{equation}
\Phi_1 =  \frac{3}{2}\frac{ {\cal M} n^2}{\Omega_o^2}  \frac{\cos(I) \sin(I)}{(1 - e_L^2)^{3/2}} \, ,
\end{equation}
\begin{equation}
\Phi_2 = \frac{3}{2}\frac{ {\cal M} n^2}{\Omega_o^2}  \frac{ \big( \cos^2(I) - \sin^2(I) \big)}{(1 - e_L^2)^{3/2}} \,,
\end{equation}
\label{eq:phi12}
\end{subequations}
with ${\cal M} = M_E/ (M+M_E)$, where $M$ is the mass of the Moon, and where $n$ is the mean motion of the Moon

\begin{equation}
n^2 = \frac{G (M_E + M)}{a_L^3} \, . \label{eq:n}
\end{equation}
Because of synchronous rotation, we set $n=\Omega_o$, although we keep them separate in our theoretical development.  Note that the square power on $e_L$ was missing in the definitions of $\Phi_1$ and $\Phi_2$ in Eqs. (A29) of DW16, a typo which we have corrected here.  Note also that the factor ${\cal M}$ in $\Phi_1$ and $\Phi_2$ was omitted in DW16 but introduced in SD18; it is a small correction, but is a better representation of the amplitude of the gravitational potential.

Equation (\ref{eq:ce1}) describes the elastic deformation of the whole Moon in response to the tidal potential from Earth.  For viscoelastic deformation, $\tilde{c}^{\,e}$ involves not only the real part of $S_{11}$ but also its imaginary part.  The expressions for $\tilde{c}_f^{\,e}$ and $\tilde{c}_s^{\,e}$ are similar to that of Equation (\ref{eq:ce1}), except
that they involve $\bar{A}_f$ and $\bar{A}_s$ and the compliances $S_{21}$ and $S_{31}$, respectively.  The perturbations in the moment of inertia tensors from external contributions are then

\begin{equation}
\tilde{c}^{\,e} =  - \bar{A} S_{11} \tilde{\phi}_m^{\,e} \, , \hspace*{0.5cm} \tilde{c}_f^{\,e} = - \bar{A}_f S_{21} \tilde{\phi}_m^{\,e} \, , \hspace*{0.5cm} \tilde{c}_s^{\,e} =   - \bar{A}_s S_{31} \tilde{\phi}_m^{\,e} \, ,
\end{equation}
where $\tilde{\phi}_m^{\,e}$ is the complex amplitude of the tidal potential from Earth acting on the whole Moon,

\begin{equation}
    \tilde{\phi}_m^{\,e} = \Phi_1  + \Phi_2 \tilde{p} \, .
    \label{eq:tildephim}
\end{equation}

The three sets of compliances ($S_{11}$, $S_{12}$, $S_{13}$), ($S_{21}$, $S_{22}$, $S_{23}$) and ($S_{31}$, $S_{32}$, $S_{33}$) describe the degree 2 deformation of the whole Moon, the fluid core, and the inner core respectively \cite[][]{mathews91a,buffett93}. For a purely elastic deformation the compliances are real.  For a viscoelastic deformation they are complex, with their imaginary part reflecting the out-of-phase response of the different regions of the Moon.  The complex compliance $S_{11} = Re[S_{11}] + i Im[S_{11}]$ captures the viscoelastic deformation of the whole Moon in response to tidal or centrifugal forcing, and is connected to $k_2$ by Equation (\ref{eq:k2}) and to the quality factor $Q$ by

\begin{equation}
 Q = \frac{Re[S_{11}]}{Im[S_{11}]} \, . \label{eq:Q}
\end{equation}

An additional deformation parameter, $\tilde{n}_{\epsilon}$, was introduced in \cite{dumberry09} to capture the change in the gravitational coupling parameter $\alpha_g$ arising as a consequence of deformation in the fluid core and solid mantle.   $\tilde{n}_{\epsilon}$ involves a fourth set of compliances ($S_{41}$, $S_{42}$, $S_{43}$) and is defined as

\begin{equation}
\tilde{n}_{\epsilon} = S_{41} ( \tilde{m}- \tilde{\phi}_m^{\,e})  + S_{42} \tilde{m}_f + S_{43} (\tilde{m}_s - \tilde{\phi}_s^{\,i}) \, .
\end{equation}

 \subsection{The gravitational torque from Earth}
 
The gravitational torque applied to the figure of the Moon by the Earth ($\boldsymbol{\Gamma_e}$) can be computed from \cite[e.g.][Equation 2]{williams01},

\begin{equation}
\boldsymbol{\Gamma_e} = -3 \frac{GM_E}{r^3} \Big( {\bf u} \times ( \boldsymbol{\cal I} \cdot {\bf u} ) \Big) \, . \label{eq:bgammae}
\end{equation}
The torque torque on the inner core ($\boldsymbol{\Gamma_{es}}$) is computed similarly, except it involves the moment of inertia tensor $\boldsymbol{{\cal I}_s}$ and the density contrast factor $\alpha_3$.  For the Cassini state of interest here, we must take the mean torque averaged over one orbital period.  Adopting our complex notation, we denote the complex amplitudes, at frequency $\omega \Omega_o$, of the external torque acting on the whole Moon and on the inner core as $\tilde{\Gamma}_e$ and $\tilde{\Gamma}_{es}$, respectively.  We take into account the triaxial figures of the whole Moon and inner core in the derivation of these torques, which involves the moment of inertia ratios $\beta$ and $\beta_s$ defined as

\begin{equation}
\beta = \frac{C - A}{B} \approx  \frac{C - A}{\bar{A}} \, , \hspace{1cm} 
\beta_s = \frac{C_s - A_s}{B_s} \approx  \frac{C_s - A_s}{\bar{A}_s} \, .
\end{equation}

An expression for the torque amplitude $\tilde{\Gamma}_e$ is presented in DW16 under the assumption of a rigid mantle and inner core,

\begin{equation}
\tilde{\Gamma}_e^{(r)}  = -i \Omega_o^2 \bar{A} \beta \left( \Phi_1 +  \Phi_2 \tilde{p} \right) -i \Omega_o^2 \bar{A}_s \beta_s \alpha_3 \Phi_2 \tilde{n}_s  \label{eq:tqr} \, ,
\end{equation}
where $\Phi_1$ and $\Phi_2$ are given in Equations (\ref{eq:phi12}).  It is straightforward to show that, in the absence of an inner core, this rigid torque is equivalent to that obtained by Equation (\ref{eq:bgammae}), due to a mass $M_E \rightarrow M_E/2$ at distance $r=a_L\sqrt{1-e_L^2}$ moving with position ${\bf u}$ as given by Equation (\ref{eq:u}) and acting on an equivalent axisymmetric Moon figure with an equatorial moment of inertia equal to $A$.  Note that the form of the contribution of the inner core to $\tilde{\Gamma}_e^{(r)}$ is slightly different from the one used in DW16, where $\beta_s$ was approximated as $e_s$.  

We must modify $\tilde{\Gamma}_e^{(r)}$ in Equation (\ref{eq:tqr}) to take into account the perturbations in the moment of inertia tensors caused by viscoelastic deformation.  In the expression of $\tilde{\Gamma}_e^{(r)}$, the part $\bar{A} \beta \tilde{p}$ captures the amplitude and orientation of the instantaneous non-spherical part of the Moon's moment of inertia on which the tidal potential from Earth acts.  Viscoelastic deformation of the whole Moon from internal origin $\tilde{c}^{\,i}$ contribute an additional part, and can be taken into account by replacing $\bar{A} \beta \tilde{p}$ with ($\bar{A} \beta \tilde{p} + \tilde{c}^{\,i})$.    

The last term on the right-hand side of Equation (\ref{eq:tqr}) represents the gravitational torque by Earth on the misaligned figure of the inner core.  For a rigid inner core, the non-spherical part of its figure which determines the amplitude of the torque is captured by $A_s \beta_s \tilde{n}_s$.  Taking into account deformation from internal origin, we must add to this the off-diagonal elements of the moment of inertia tensor captured by $\tilde{c}_s^{\,i}$ \cite[][]{dumberry09}.  As above, this is done by substituting $A_s \beta_s \tilde{n}_s$ with ($A_s \beta_s \tilde{n}_s + \tilde{c}_s^{\,i}$).

The viscoelastic deformation associated with the external gravitational potential from Earth further contributes to a change in the torque.  For a planetary body in synchronous rotation, and in the limit of small misalignment angles, \cite{baland16} and \cite{coyette16} have shown that the moment of inertia difference $(C-A)$ involved in the torque is modified by elastic deformation to $(C-A) - k_2 q_r M R^2$, where $q_r = (M_E/M) \cdot (R/a_L)^3$.  Let us denote $\beta'$ the modified $\beta$ factor that participates in the torque.  Using Equations (\ref{eq:k2}) and (\ref{eq:n}), $\beta'$ is then equal to

\begin{equation}
\beta' = \beta - 3  \frac{n^2}{\Omega_o^2} {\cal M} \cdot Re[S_{11}] \, .
\end{equation}
Elastic deformations of the inner core lead to a similar modification of the moment of inertia difference $(C_s-A_s)$. By analogy with the whole Moon, we denote $\beta_s'$ the modified $\beta_s$ factor caused by the external potential,

\begin{equation}
\beta_s' = \beta_s - 3  \frac{n^2}{\Omega_o^2} {\cal M} \cdot Re[S_{31}] \, .
\end{equation}
The torque on the whole Moon, taking into account elastic deformation, is then 

\begin{align}
\tilde{\Gamma}_e & = -i \Omega_o^2 \left[ \bar{A} \beta' \tilde{\phi}_m^{\,e} + \bar{A}_s \beta'_s \tilde{\phi}_s^{\,e} + \Phi_2(\tilde{c}^{\,i} + \alpha_3 \tilde{c}_s^{\,i} ) \right] \nonumber  \\
& = -i \Omega_o^2 \left[ \bar{A} \Big(\beta - 3  \frac{n^2}{\Omega_o^2} {\cal M} \cdot Re[S_{11}] \Big) \tilde{\phi}_m^{\,e} + \bar{A}_s \Big(\beta_s- 3  \frac{n^2}{\Omega_o^2} {\cal M} \cdot Re[S_{31}] \Big) \tilde{\phi}_s^{\,e} + \Phi_2 \big( \tilde{c}^{\,i} + \alpha_3 \tilde{c}_s^{\,i} \big) \right]  \label{eq:tqe} \, ,
\end{align}
where $\tilde{\phi}_m^{\,e}$ is given by Equation (\ref{eq:tildephim}) and $\tilde{\phi}_s^{\,e}$ is defined by

 \begin{equation}  
 \tilde{\phi}_s^{\,e} = \alpha_3 \Phi_2 \tilde{n}_s \, . 
\end{equation}
 
When dissipation is present the tidal deformation is no longer in phase with the imposed gravitational potential.  In our system, this involves the imaginary part of the compliances $S_{11}$ and $S_{31}$.   This gives rise to a tidal torque, but we consider this torque separately in the next section.  

The gravitational torque on a rigid inner core by the Earth, expressed by Eqs. A34-A35 of DW16, is

\begin{align}
\tilde{\Gamma}_{es}^{(r)} & = -i \Omega_o^2 \bar{A}_s \alpha_3 \beta_s  \left( \Phi_1 + \Phi_2 \tilde{p} + \Phi_2 \tilde{n}_s  \right) \, , \nonumber \\  
& = -i \Omega_o^2 \bar{A}_s  \beta_s \left( \alpha_3  \tilde{\phi}_m^{\,e} + \tilde{\phi}_s^{\,e}  \right) \, . \label{eq:tqser}
\end{align}
To take into account viscoelastic deformation of the inner core from internal potentials, we proceed as above and replace $A_s \beta_s \tilde{n}_s$ with ($A_s \beta_s \tilde{n}_s + \tilde{c}_s^{\,i}$) in Equation (\ref{eq:tqser}). Under the influence of the external potential, elastic deformation leads to a modification of $\beta_s$ to  $\beta'_s$. The gravitational torque on a viscoelastic inner core by the Earth is then  

\begin{align}
\tilde{\Gamma}_{es} & = -i \Omega_o^2 \left[ \bar{A}_s \beta'_s \left( \alpha_3 \tilde{\phi}_m^{\,e} + \tilde{\phi}_s^{\,e} \right) +  \alpha_3 \Phi_2  \tilde{c}_s^{\,i} \right] \, ,\nonumber\\
& = -i \Omega_o^2 \left[ \bar{A}_s \Big(  \beta_s - 3  \frac{n^2}{\Omega_o^2} {\cal M} \cdot Re[S_{31}] \Big) \left( \alpha_3 \tilde{\phi}_m^{\,e} + \tilde{\phi}_s^{\,e} \right) +  \alpha_3 \Phi_2  \tilde{c}_s^{\,i} \right] \, .  \label{eq:tqse} 
\end{align}

\subsection{The torque from tidal dissipation}

Dissipation results in a time delay $\Delta t$ between the imposed gravitational potential from Earth and the tidal resonse of the Moon. The amplitude of the time delay depends on the frequency (the mean motion $n$ for the Cassini state) and the energy lost through dissipation within the Moon, captured by the quality factor $Q$, and can be modelled as $\Delta t = 1 / (nQ)$.  Because of the misalignment between the tidal maximum and the imposed potential, an additional torque acts on the Moon which we refer to as the tidal torque, and is labelled as ${\bf \Gamma_t}$ in Equation (\ref{eq:dHdt}).  A similar tidal torque acts on the inner core (${\bf \Gamma_{ts}}$).

One component of the tidal torque is in the opposite direction of the rotation vector ${\bf \Omega}$; this is the component that has led to tidal locking of the Moon by Earth.  The tidal torque also has a component in the direction of the normal to the orbital plane ($\mitbf{\hat{e}_3^I}$). This is the component of the tidal torque which induces a phase lead angle $\phi_p$. 

Derivations of the torque associated with tidal dissipation for a planetary body in a 1:1 spin orbit resonance, as appropriate for the Moon, can be found in many studies.  Averaged over one orbit, the component of this torque along $\mitbf{\hat{e}_3^I}$ is given by \cite[e.g.][Equations 1-2]{levrard07} 

\begin{equation}
{\bf \Gamma_t} = 3 \frac{k_2}{Q} \frac{G M_E^2 R^5}{a_L^6} \left( f_1(e_L) - \frac{f_2(e_L)}{2n} {\bf \Omega} \cdot  \mitbf{\hat{e}_3^I} \right) \mitbf{\hat{e}_3^I} \, , \label{eq:td1}
\end{equation}
where the functions of the eccentricities $f_1(e_L)$ and $f_2(e_L)$ are given by

\begin{subequations}
\begin{align}
 f_1(e_L) & = \frac{1 + \frac{15}{2} e_L^2 + \frac{45}{8} e_L^4}{(1-e_L^2)^6} \, ,\\ 
f_2(e_L) & = \frac{1 + 3 e_L^2 + \frac{3}{8} e_L^4}{(1-e_L^2)^{9/2}} \, .
\end{align}
\label{eq:f1f2}
\end{subequations}
Written in terms of the complex compliance $S_{11}$ introduced in section 2.3, and using the definition of the mean motion $n$,  Equation (\ref{eq:td1}) becomes

\begin{equation}
{\bf \Gamma_t} = 9 \cdot Im[S_{11}] \cdot \bar{A} {\cal M}^2  \frac{n^4}{\Omega_o^2} \left( f_1(e) -\frac{ f_2(e)}{2 n} {\bf \Omega} \cdot \mitbf{\hat{e}_3^I}   \right) \mitbf{\hat{e}_3^I}  \, .\label{eq:td2}
\end{equation}

We now project this torque onto the equatorial components of our Moon-attached $\mitbf{\hat{e}^p}$-frame. Viewed from the $\mitbf{\hat{e}^p}$-frame, the orientation of $\mitbf{\hat{e}_3^I}$ is rotating at frequency $\omega \Omega_o$.  In our complex plane notation, the mean tidal torque is then periodic and proportional to $\exp[ i \omega \Omega_o t]$.  At $t=0$, $\mitbf{\hat{e}_3^I}$ points toward $- \mitbf{\hat{e}_1^p}$, in other words, in the negative real direction of the complex plane.  The angle between $\mitbf{\hat{e}_3^I}$ and $\mitbf{\hat{e}_3^p}$ is $(I+\theta_p)$, so the projection of the tidal torque onto the complex plane is

\begin{equation}
\tilde{\Gamma}_t = - \Omega_o^2 \bar{A} \cdot Im[S_{11}] \cdot  \Phi^t \label{eq:td3}
\end{equation}
where 

\begin{equation}
 \Phi^t  =  9 {\cal M}^2  \frac{n^4}{\Omega_o^4}  \sin(I+\theta_p) \left( f_1(e) - \frac{ f_2(e)}{2} \frac{\Omega_o}{n} \cos (I + \theta_p) \right) \, , \label{eq:Phit}
\end{equation}
and where, since $\theta_m \ll \theta_p$, we have used the following approximation 

\begin{equation}
{\bf \Omega} \cdot \mitbf{\hat{e}_3^I} = \Omega_o \cos(I +\theta_p +\theta_m) \approx \Omega_o \cos(I +\theta_p) \, .
\end{equation}
For the small tilt angles approximation, $\theta_p \approx \tilde{p} \ll 1$, we can write the tidal torque as

\begin{equation}
\tilde{\Gamma}_t = - \Omega_o^2 \bar{A} \cdot Im[S_{11}] \cdot \left( \Phi_1^t + \Phi_2^t \tilde{p}  \right) \label{eq:td4}
\end{equation}
where 

\begin{subequations}
\begin{align}
 \Phi_1^t  &=  9 {\cal M}^2 \frac{n^4}{\Omega_o^4} \left(f_1(e) \sin I  - \frac{f_2(e)}{2} \frac{\Omega_o}{n}   \cos I \, \sin I \right) \, , \\
 \Phi_2^t &=  9 {\cal M}^2 \frac{n^4}{\Omega_o^4} \left(f_1(e) \cos I  - \frac{f_2(e)}{2} \frac{\Omega_o}{n}   (\cos^2 I \, - \sin^2 I ) \right) \, .
\end{align}
\end{subequations}
We note again that, because of synchronous rotation, we set $n=\Omega_o$ in our study.

For a Moon with an inner core rotating at an angular velocity ${\bf \Omega_s}$ different to that of the mantle, the tidal dissipation torque on the whole of the Moon in Equation (\ref{eq:td2}) can instead be expressed as 

\begin{align}
{\bf \Gamma_t} & = 9 \cdot Im[S_{11}] \cdot \bar{A} {\cal M}^2  \frac{n^4}{\Omega_o^2} \left( f_1(e) -\frac{ f_2(e)}{2n} {\bf \Omega} \cdot  \mitbf{\hat{e}_3^I} \right) \mitbf{\hat{e}_3^I} \nonumber\\
&+  9 \cdot Im[S_{31}] \cdot \bar{A}_s  \alpha_3 {\cal M}^2  \frac{n^4}{\Omega_o^2} \frac{ f_2(e)}{2n}  \Big({\bf \Omega} \cdot  \mitbf{\hat{e}_3^I} -  {\bf \Omega_s} \cdot  \mitbf{\hat{e}_3^I}  \Big) \mitbf{\hat{e}_3^I}\, . 
\label{eq:td5}
\end{align}
Under the small angle approximation, the presence of the inner core modifies the torque on the whole Moon of Equation (\ref{eq:td4}) to

\begin{equation}
 \tilde{\Gamma}_t = - \Omega_o^2 \bar{A} \cdot Im[S_{11}] \cdot \left( \Phi_1^t  + \Phi_2^t \tilde{p}  \right) - \Omega_o^2 \bar{A}_s \cdot Im[S_{31}] \cdot \alpha_3 \Phi_3^t \tilde{m}_s \, , \label{eq:td6}
\end{equation}
with

\begin{equation}
 \Phi_3^t  =  \frac{9}{2}  \frac{n^3}{\Omega_o^3} f_2(e) {\cal M}^2  \sin^2 I \, .
\end{equation}

The tidal torque acting on the inner core alone can be derived from Equation (\ref{eq:td2}) and by substituting with the appropriate quantities for the inner core,

\begin{equation}
{\bf \Gamma_{ts}} = 9 \, \bar{A}_s \alpha_3 {\cal M}^2  \frac{n^4}{\Omega_o^2} \left(Im[S_{31}] \cdot  f_1(e) -Im[S_{33}] \cdot \frac{ f_2(e)}{2 n} {\bf \Omega_s} \cdot {\bf \hat{e}_3^I}  \right) {\bf \hat{e}_3^I} \label{eq:tqdsic} \, .
\end{equation}
Written in our complex notation, and under the small angle approximation, the tidal torque on the inner core is 
 
\begin{equation}
\tilde{\Gamma}_{ts} =  - \Omega_o^2 \bar{A}_s \alpha_3 \Big( Im[S_{31}] \cdot \left( \Phi_1^t + \Phi_2^t \tilde{p} \right)  + Im[S_{33}] \cdot \Phi_3^t \tilde{m}_s  \Big) \, . \label{eq:tqdsic2} 
\end{equation}

\subsection{Internal torques}

The torques on the inner core ($\boldsymbol{\Gamma_{icb}}$) and on the fluid core ($\boldsymbol{\Gamma_{cmb}}$) are caused by viscous tractions acting on the inner core and on the mantle side of the CMB, respectively.  We follow previous authors \cite[e.g.][]{mathews05} and use a parameterization for these torques involving a product of dimensionless complex coupling constants $K_{icb}$ and $K_{cmb}$ and the differential angular velocities at each boundary.   Using our complex notation, these torques are written as

\begin{subequations}
\begin{equation}
    \tilde{\Gamma}_{icb} = i \Omega_o^2 \bar{A}_s K_{icb} (\tilde{m}_f - \tilde{m}_s) \, ,\label{eq:tqicb}
\end{equation}
\begin{equation}
    \tilde{\Gamma}_{cmb} = i \Omega_o^2 \bar{A}_f K_{cmb} \, \tilde{m}_f \, .\label{eq:tqcmb}
\end{equation}

Expressions for $K_{icb}$ and $K_{cmb}$ are given by \cite{mathews05}, but these are based on the assumption that the flow remains laminar.  Because of the large differential velocity between the mantle and fluid core of the Moon, the flow close to the CMB is most likely turbulent \cite[][]{toomre66,yoder81,williams01,cebron19}. Likewise, flow near the ICB is also likely turbulent given the expected large misalignment between the rotation vectors of the fluid and solid cores (DW16, SD18).  An estimate of the magnitude of this turbulent viscous coupling is inferred from LLR observations, through a dissipation parameter $K$, and reported in terms of the ratio $K/C$, where $C$ is the polar moment of inertia of the Moon \cite[][]{williams01,williams15}. This inference is based on a lunar model that does not have an inner core, in which case $K_{icb}=0$, and $K_{cmb}$ is connected to $K/C$ by

\begin{equation}
    Im[K_{cmb}] = - \bigg( \frac{K}{C} \bigg) \bigg( \frac{\bar{A}}{\bar{A}_f} \bigg) \bigg( \frac{1}{\Omega_o} \bigg) \, .\label{eq:kcmbt2}
\end{equation}
\end{subequations}

The expression for the pressure torque and the gravitational torque from the rest of the Moon acting on the inner core ($\boldsymbol{\Gamma_s}$) presented in DW16 neglects viscoelastic deformation.  A derivation of this torque which includes the effects of deformation is presented in \cite{dumberry09}, and is given by

\begin{equation}
    \tilde{\Gamma}_s = i \Omega_o^2 \bar{A}_s \bigg( -e_s \alpha_1 (\tilde{m} + \tilde{m}_f) + e_s \alpha_2 \tilde{n}_s + \alpha_2 \frac{\tilde{c}_s}{\bar{A}_s} + e_s \alpha_3 \alpha_g \tilde{n}_{\epsilon} \bigg) \, .    \label{eq:tqs}
\end{equation}

\subsection{The linear system of equations}

Since each of the rotational variables are proportional to $\exp [i \omega \Omega_o t]$, their time-derivatives can be replaced by $i \omega \Omega_o$.   A solution of the system of Equations (\ref{eq:sys}) can be obtained in the frequency domain at the specific forcing frequency $\omega = -1 - \delta \omega$ associated with the Cassini state.  Under the assumption that the dynamical ellipticities $e$, $e_f$, $e_s$ are all much smaller than unity, and similarly for all of the compliances, the left-hand side of each of the five equations of our rotational model can be linearized in terms of the rotational variables.  The system of Equations (\ref{eq:sys}) can be written as

\begin{subequations}
\begin{equation}
 (\omega - e)\tilde{m} + (1 +\omega) \Bigg[ \frac{\bar{A}_f}{\bar{A}} \tilde{m}_f + \frac{\bar{A}_s}{\bar{A}}  \tilde{m}_s + \alpha_3 e_s \frac{\bar{A}_s}{\bar{A}} \tilde{n}_s + \frac{\tilde{c}}{\bar{A}} \Bigg] = \frac{1}{i \Omega_o^2 \bar{A}} \Big(\tilde{\Gamma}_e  + \tilde{\Gamma}_t \Big) \, , \label{eq:am1}
 \end{equation}
 
 \begin{equation}
 \omega \tilde{m} + \left( 1 + \omega  + e_f  \right) \tilde{m}_f -  \omega \alpha_1 e_s \frac{\bar{A}_s}{\bar{A}_f} \tilde{n}_s + \omega \frac{\tilde{c}_f}{\bar{A}_f} =   \frac{1}{i \Omega_o^2 \bar{A}_f} \Big(- \tilde{\Gamma}_{cmb}  - \tilde{\Gamma}_{icb} \Big) \, , \label{eq:af1}
 \end{equation}
 
 \begin{equation}
( \omega - e_s) \tilde{m} +  \left(1+\omega\right) \Bigg[  \tilde{m}_s 
+ e_s \tilde{n}_s + \frac{\tilde{c}_s}{\bar{A}_s} \Bigg]= \frac{1}{i \Omega_o^2 \bar{A}_s} \Big( \tilde{\Gamma}_{es}  + \tilde{\Gamma}_{ts} + \tilde{\Gamma}_{s} + \tilde{\Gamma}_{icb} \Big) \, , \label{eq:as1}  
 \end{equation}
 
 \begin{equation}
 \tilde{m}_s  + \omega  \tilde{n}_s  = 0 \, , \label{eq:msns}
 \end{equation}
 
 \begin{equation}
\tilde{m} + (1+\omega) \tilde{p} = 0 \, .
 \end{equation}
 \label{eq:sys2}
 \end{subequations}

Substituting the expressions for each of the torques on the right-hand sides defined in Equations (\ref{eq:tqe}), (\ref{eq:tqse}), (\ref{eq:td6}), (\ref{eq:tqdsic2}), (\ref{eq:tqicb}), (\ref{eq:tqcmb}) and (\ref{eq:tqs}), and using the definition of the compliances in section 2.3, Equations (\ref{eq:sys2}) can be written as a linear system with solution vector $\textbf{x} = [\tilde{m}, \tilde{m}_f, \tilde{m}_s, \tilde{n}_s, \tilde{p}]^T$ as follows,

\begin{equation}
\boldsymbol{\mathsf{M}} \cdot \textbf{x} = \textbf{y} \, , \label{eq:5by5}
\end{equation}
where the forcing vector $\textbf{y}$ and the elements of the 5-by-5 matrix $\boldsymbol{\mathsf{M}}$ are given in Appendix A.  Note that the ordering of the vector $\textbf{x}$ is different than that in DW16.  

In the absence of an inner core, the rotational dynamics is fully described by the three Equations (\ref{eq:dHdt}), (\ref{eq:dHfdt}) and (\ref{eq:de3dt}).  The truncated linear system is written as

\begin{equation}
\boldsymbol{\mathsf{M}}' \cdot \textbf{x}' = \textbf{y}' \, ,\label{eq:3by3}
\end{equation}
where $\textbf{x}' = [\tilde{m}, \tilde{m}_f, \tilde{p}]^T$.  The elements of $\textbf{y}'$ and $\boldsymbol{\mathsf{M}}'$ are given in Appendix A.

The last equation in each of these two linear systems is the kinematic relation of Equation (\ref{eq:de3dt}) which, under the small angle approximation, gives a simple relationship between $\tilde{p}$ and $\tilde{m}$ (DW16),

\begin{equation}
\tilde{p} = - \frac{ \tilde{m}}{1 + \omega} = \frac{ \tilde{m}}{\delta \omega}\, .
\end{equation}
Since the Poincar\'e number, $\delta \omega$,  is much smaller than unity, $|\tilde{p}| \gg | \tilde{m} |$.  In other words, although we take into account the offset between the rotation vector and the figure axis of the mantle, this offset is much smaller than the tilt angle of the mantle figure axis of $1.543^\circ$ with respect to the ecliptic.  Indeed, in describing the Cassini state of the Moon, it is often the rotation axis of the Moon which is referred to as being tilted by $1.543^\circ$, with the underlying assumption that the rotation and figure axes are perfectly aligned.  Likewise, the departure from an exact Cassini state is often described in terms of the phase lead angle of 0.27 arcsec of the rotation vector of the Moon.  But to be precise, LLR observations track the motion of the solid mantle, and thus the offset of its figure axis.  Hence, in our system, the observed phase lead angle $\phi_p$ is connected to the imaginary part of $\tilde{p}$.  Because the Cassini plane rotates in a retrograde sense, at $t=0$ a phase lead corresponds to a negative imaginary component.  Hence, under the approximation of small angles, $\phi_p = - Im[\tilde{p}]$.  

\subsection{Computation of the compliances}

The method for computing the compliances $S_{ij}$ is presented in detail in many places \cite[e.g.][]{alterman59,dehant15} and will not be repeated here.  We follow closely the approach presented in \cite{dumberry04} and \cite{dumberry08b}.  The method consists of integrating a system of six coupled ordinary differential equations in radius from a small radius near the centre to the lunar surface.  Two of the six computed variables are the vertical and lateral displacements at every point inside the planet.  These displacements are then used to compute the perturbations in the moments of inertia, from which the compliances $S_{ij}$ are finally calculated. 

Each of the four solid regions in our model (crust, mantle, LVZ and inner core) are assumed to have the viscoelastic rheology of a Maxwell solid.  That is, we assume that the Lam\'e parameter $\lambda$ and the shear modulus $\mu$ are frequency dependent and we use a constitutive relation

\begin{equation}
\boldsymbol{\mathsf{T}}= \lambda(\omega') \boldsymbol{\mathsf{I}} ( \boldsymbol{\nabla} \cdot \textbf{u} ) + \mu(\omega') (\boldsymbol{\nabla} \textbf{u} + (\boldsymbol{\nabla} \textbf{u})^T) ,
\end{equation}
where $\boldsymbol{\mathsf{T}}$ is the incremental Lagrangian-Cauchy stress tensor and the vector $\textbf{u} = \textbf{u}(\textbf{r})$ describes the displacement of material particles specified by the position vector $\textbf{r}$. The frequency-dependent $\lambda(\omega')$ and $\mu(\omega')$ are specified by \cite[e.g.][]{wu82,koot11}

\begin{subequations}
\begin{equation}
    \lambda(\omega') = \frac{( i \omega' \lambda_o + \frac{\kappa}{\eta} \mu_o )}{(i \omega' + \frac{1}{\eta} \mu_o)} \, , \label{eq:lambdap}
\end{equation}
\begin{equation}
    \mu(\omega') = \frac{ i \omega' \mu_o}{( i \omega' + \frac{1}{\eta} \mu_o )} \, ,\label{eq:mup}
\end{equation}

\noindent where $\omega' = \omega \Omega_o$, $\eta$ is the viscosity, $\kappa$ is the bulk modulus 

\begin{equation}
    \kappa = \lambda_o + \frac{2}{3} \mu_o \, ,
\end{equation}
\label{eq:lambdamu}
\end{subequations}
and  $\lambda_o$ and $\mu_o$ are the Lam\'e parameter and shear modulus in the elastic limit.   This enables us to calculate the viscoelastic response of the Moon to a forcing applied with a frequency $\omega'$.  We assume the viscosity to be uniform within a solid region.  With such a rheology, $\lambda(\omega')$ and $\mu(\omega')$ have both a real and complex part.  Hence, the computed compliances $S_{ij}$ are also complex.

Modelling the lunar mantle as a Maxwell solid is not fully compatible with the frequency dependence of the tidal dissipation which is inferred by LLR; indeed more sophisticated rheologies have been proposed \cite[e.g.][]{williams01,williams14,williams15}.  However, here we are focused on the dissipation occurring at a single frequency -- that associated with the Cassini state.  Our use of a Maxwell solid model is not intended to capture the true rheology of the LVZ and the mantle above.  Rather, this choice is motivated by its simplicity.  Likewise, our use of a Maxwell rheology for the inner core is motivated by its simplicity.  Our primary goal is to show that it is possible to capture tidal dissipation in the solid part of the Moon in our rotational model and to investigate whether deformation within the inner core may contribute to the observed dissipation.  More sophisticated rheologies for the mantle layers and the inner core can be incorporated in future studies.  
  
\section{Interior Moon models}

\begin{table}
\begin{tabular}{ll}
\hline
Moon Parameter & Numerical value  \\ \hline
rotation rate, $\Omega_o=n$ &  $2.6617 \times 10^{-6}$ s$^{-1}$ \\
orbit precession rate, $\Omega_p$ & $2\pi /18.6$ yr$^{-1}$ \\
Poincar\'e number, $\delta \omega = {\Omega_p}/{\Omega_o}$ & $4.022 \times 10^{-3}$ \\
mean planetary radius, $R$ & $1737.151$ km\\
mass, $M$ & $7.3463 \times 10^{22}$ kg \\
mean density, $\bar{\rho}$ & $3345.56$ kg m$^{-3}$ \\
moment of inertia of solid Moon, $I_{sm}$ & $0.393112 \cdot M R^2$ \\$J_2$  & $2.03504 \times 10^{-4}$ \\
$C_{22}$ & $2.24482 \times 10^{-5}$ \\
polar surface flattening, $\epsilon_r$ & $1.2899 \times 10^{-3}$\\
equatorial surface flattening, $\xi_r$ & $2.4346 \times 10^{-4}$\\
\hline
\end{tabular}
\caption{\label{table:parameters} Reference parameters for the Moon.  The values of $R$, $M$, $\bar{\rho}$, $I_{sm}$, $J_2$ and $C_{22}$ are taken from \cite{williams14}. The values for the unnormalized potential coefficients  $J_2$ and $C_{22}$ include the permanent tide from synchronous rotation around Earth, and are obtained after multiplying the values reported in \cite{williams14} by a factor 1.000978 to take into account our definition of the mean radius as $R=1737.151$ km instead of $1738$ km used in the GRAIL-derived gravity field. $\epsilon_r$ and $\xi_r$ are taken from \cite{araki09} and converted to our choice of normalization.}
\end{table}

Each model of the interior density structure of the Moon is constrained by: the lunar mass $M = ( 4 \pi / 3) \bar{\rho} R^3$, where $\bar{\rho} = 3345.56$ kg m$^{-3}$ is the mean density and $R = 1737.151$ km is the mean radius; and by the moment of inertia of the solid Moon $I_{sm} = 0.393112  M R^2$ \cite[][]{williams14}, comprised here of the LVZ, mantle and crust.  The two following constraints must be satisfied,

\begin{subequations}
\begin{align}
\bar{\rho} R^3 &= \rho_s r_s^3 + \rho_f \left(r_f^3 - r_s^3\right) + \rho_l \left(r_l^3 - r_f^3\right) + \rho_m \left(r_m^3 - r_l^3\right) + \rho_c \left(R^3 - r_m^3\right) \, , \label{eq:rho}\\
I_{sm} & = \frac{8 \pi}{15} \Big( \rho_l \left(r_l^5 - r_f^5\right) + \rho_m \left(r_m^5 - r_l^5\right) + \rho_c \left(R^5 - r_m^5\right) \Big) \, . \label{eq:Ism}
\end{align}
\label{eq:rhoIsm}
\end{subequations}

For all interior models in our study we use a fixed crustal layer with a thickness of 38.5 km and a density of $2550$ kg m$^{-3}$ \cite[][]{wieczorek13}.  When present, the inner core density is fixed at 7700 kg m$^{-3}$ \cite[][]{matsuyama16}.  The outer radius of the LVZ is fixed at $r_l=550$ km; this is consistent with the results of \cite{matsumoto15} and \cite{harada16}.  We further assume no density contrast between the mantle and the LVZ (i.e. $\rho_l=\rho_m$).  The radii of the fluid core and inner core are varied over a range of acceptable values, and for each set of $r_f$ and $r_s$, the density of the mantle (and LVZ) is set by Equation (\ref{eq:Ism}) and the density of the fluid core is then determined by Equation (\ref{eq:rho}). 

Once the radial density structure is specified, we determine the polar and equatorial flattenings at each boundary.  These are constrained by the degree 2 gravitational potential coefficients $J_2$ and $C_{22}$ and the observed surface flattenings $\epsilon_r$ and $\xi_r$.  We assume that the ICB and CMB are both at hydrostatic equilibrium with the imposed gravitational potential from the LVZ, mantle and crust.  With the assumption of $\rho_l=\rho_m$, the flattenings at the LVZ-mantle boundary ($\epsilon_l$ and $\xi_l$) do not contribute to the gravitational potential and the procedure to determine the polar and equatorial flattenings at the ICB, CMB and crust-mantle boundary is equivalent as that detailed in section 3.1 of SD18. With the complete triaxial shape of the Moon specified, we can then compute the dynamical ellipticities $e_s$, $e_f$ and $e$, as well as the moment of inertia ratios $\beta_s$ and $\beta$.  The numerical values of the parameters used to specify our Moon models are listed in Table \ref{table:parameters}.

The third and last step is to compute the set of compliances for each interior model.  To do so, we must first specify the Lam\'e parameters in the elastic limit ($\lambda_o$ and $\mu_o$) within each region.  These are related to the density ($\rho$) and the compressional ($V_p$) and shear ($V_s$) seismic wave velocities by

\begin{equation}
\mu_o = \rho V_s^2 \, \hspace*{1cm}  \lambda_o = \rho V_p^2 - 2 \mu_o \, .
\end{equation}
In the fluid core, where $V_s=0$, $\mu_o=0$.  We assume $V_s$ and $V_p$ are uniform within each layer and their values are specified in Table \ref{table:seis}; they are broadly consistent with the lunar seismic velocity profiles derived from the Apollo Passive Seismic Experiment (APSE) \cite[e.g.][]{garcia11,weber11,matsumoto15}.  The frequency dependent Lam\'e parameters are then computed according to Equation (\ref{eq:lambdamu}), and depend on the choice of viscosity adopted for each solid region.  We present results for a range of viscosities for the LVZ and inner core.  However, to keep our results tractable and to keep the focus on the effects of lower mantle and inner core deformation, we set the viscosities of both the mantle and crust to a fixed value of $1 \times 10^{20}$ Pa s.  This is broadly consistent with published estimates based on the observed topography, history of lunar volcanism and thermal and chemical evolution \cite[e.g.][]{hess95,zhong00}.

\begin{table}[]
\begin{tabular}{llllll}
\hline
Seismic parameter & Crust & Mantle & LVZ & Fluid core & Inner core \\ \hline
$V_p$ (m s$^{-1}$) & 4000 & 8000 & 7500 & 4000 & 4200 \\
$V_s$ (m s$^{-1}$) & 2000 & 4500 & 3500 & 0 & 2200 \\
$\rho$ (kg m$^{-3}$) & 2550 & calculated & calculated  & calculated & 7700 \\
\hline
\end{tabular}
\caption{\label{table:seis} Seismological parameters used in our calculations.  $V_p$ and $V_s$ are, respectively, the compressional and shear seismic velocities.  The density ($\rho$) for the mantle, LVZ and fluid core are model dependent; they depend on the choice of CMB and ICB radii and constrained to match $I_{sm}$ and $\bar{\rho}$ as determined by Equation \ref{eq:rhoIsm}.}
\end{table}

\section{Results I: Tidal deformation and viscous friction at the CMB in the absence of an inner core}

We first present results based on a set of interior models that do not have an inner core.   This is to ensure that our rotational model is consistent with previously published models of lunar dissipation.  More specifically, that our model recovers the observed phase lead of $\phi_p=0.27$ arcsec ahead of the Cassini plane when the relative contributions from tidal dissipation and viscous friction at the CMB estimated in previous studies are used as inputs.  A secondary objective is to better illuminate how each of these two contributions affects $\phi_p$.

\subsection{Viscoelastic deformation of the Moon}

We begin by investigating how the viscosity of the LVZ influences the viscoelastic deformation of the whole Moon.  We do so over a range of plausible fluid core radii, spanning between 340 km to 420 km \cite[consistent with previous estimates, e.g.][]{weber11,garcia11,matsumoto15,matsuyama16}, and LVZ viscosities ($\eta_{lvz}$) from $10^{14}$ to $10^{18}$ Pa s.  Note that for a core radius smaller than $\sim$305 km, the fluid core density in our model exceeds that of solid iron in the face-centered cubic phase (7700 kg$\cdot$m$^{-3}$) \cite[][]{tsujino13}; this sets a lower bound for our choice of CMB radius. 

\begin{figure}
\begin{center}
    \includegraphics[height=12cm]{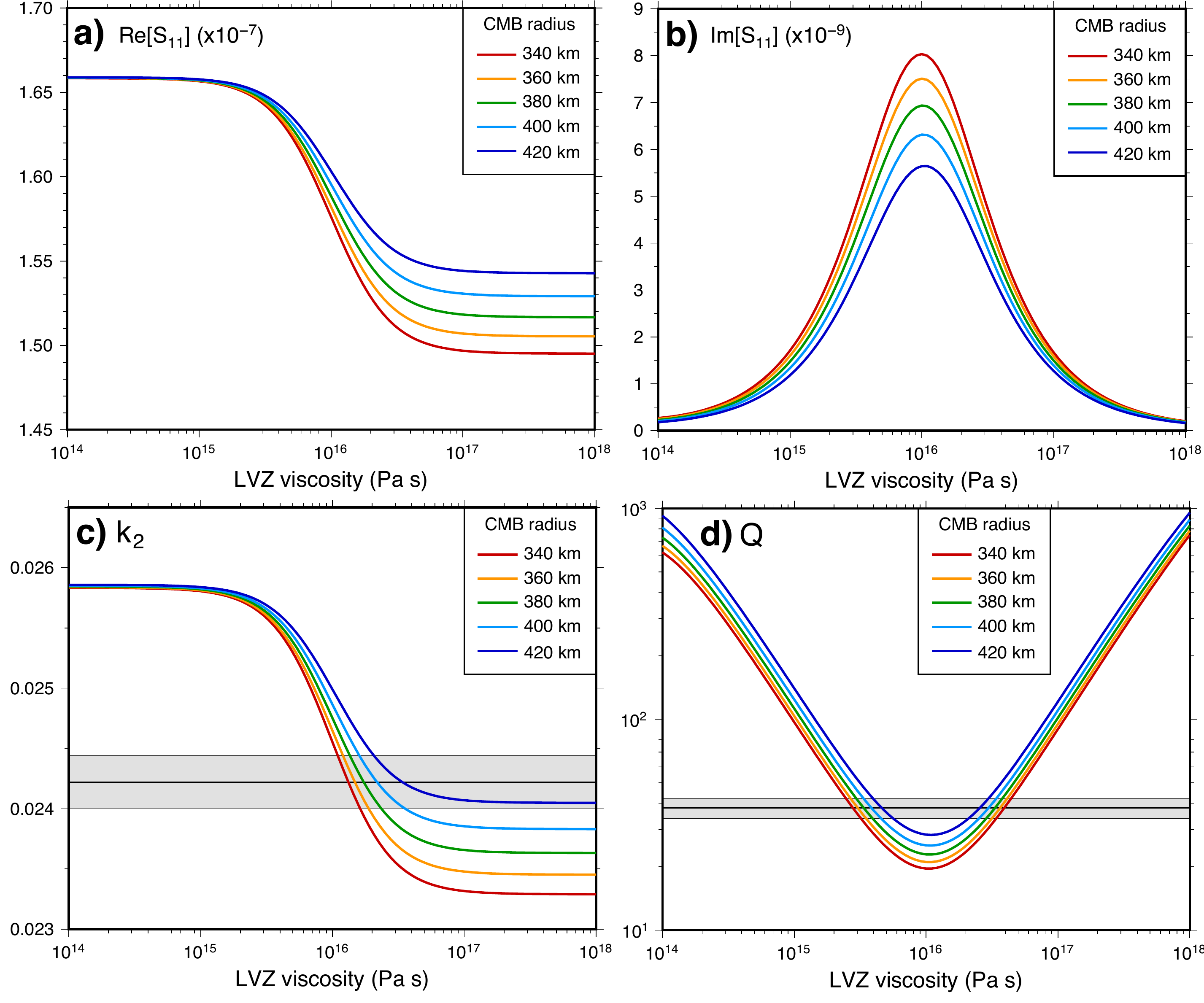} 
    \caption{\label{fig:S11_nosic} a) Real and b) Imaginary parts of the compliance $S_{11}$, c) $k_2$ and d) $Q$ as a function of LVZ viscosity for different choices of CMB radius. The grey shaded area shows the range of the estimated values of $k_2 = 0.02422\pm0.00022$ and $Q=38 \pm4$ of the Moon.  In all cases shown, the LVZ outer radius is 550 km and there is no inner core.}
\end{center}
\end{figure}

For each combination of $\eta_{lvz}$ and $r_f$ we compute the four compliances ($S_{11}$, $S_{12}$, $S_{21}$, $S_{22}$) that enter the truncated rotational model of Equation (\ref{eq:3by3}).  To be specific, these are computed at a frequency of $\omega \Omega_o = - 2 \pi/ 27.212$ day$^{-1}$, which we refer to as the monthly frequency.  We focus on $S_{11}$, which describes  the global deformation resulting from changes in both the tidal and centrifugal potentials.  Figures \ref{fig:S11_nosic}a,b show the compliance $S_{11}$ as a function of $\eta_{lvz}$ for different choices of fluid core radius.  The behaviour of $S_{11}$ is characteristic of a Maxwell rheology.   The LVZ behaves as a fluid and as an elastic solid for, respectively, low and high values of $\eta_{lvz}$.  The imaginary part of $S_{11}$ approaches zero in both the fluid ($\eta_{lvz} \rightarrow 0$) and elastic ($\eta_{lvz} \rightarrow \infty$) limits.  The transition from fluid to elastic is centred around a LVZ viscosity of approximately $10^{16}$ Pa s, which marks the point where the imaginary part of $S_{11}$ is maximum. The upper bound of the real part of $S_{11}$ at low $\eta_{lvz}$ is set by the fact that both the mantle and crust, with their relatively large viscosities, remain in the elastic limit.  

Figures \ref{fig:S11_nosic}c,d show how  $k_2$ and the monthly-$Q$, linked to $S_{11}$ by Equation (\ref{eq:Q}), vary as a function of $\eta_{lvz}$ and $r_f$.   The behaviour of $k_2$ tracks that of the real component of $S_{11}$.  The changes in $Q$ as a function of $\eta_{lvz}$ mostly reflect the changes in the imaginary part of $S_{11}$.  Energy dissipation is maximized when the value of $Q$ is at its lowest. This occurs when $Im[S_{11}]$ is largest and when the LVZ viscosity is approximately $10^{16}$ Pa s. 

Estimates of $k_2$ for the Moon can be inferred from LLR observations \cite[][]{williams01} or more directly measured using tracking data from satellites in lunar orbit \cite[][]{goossens08}.  The latest estimate of $k_2$ is based on data from the GRAIL satellite mission and is $0.02422 \pm 0.00022$ (when using a mean Moon radius $R=1737.151$ km) \cite[][]{williams14}.   Estimates of $Q$ have been obtained from measurements of the time delay between tidal forcing and the lunar response based on LLR observations \cite[][]{williams01}.  The most recent estimate of $Q$ at a monthly period is $38 \pm 4$ \cite[][]{williams15}.  Figures \ref{fig:S11_nosic}c,d show that for a LVZ viscosity in the range of $2-4 \times 10^{16}$ Pa s, both the observed $k_2$ and $Q$ can be matched. This is consistent with the findings of \cite{harada14,harada16}.

\subsection{The Phase Lead of the Cassini State}

For each combination of $\eta_{lvz}$ and $r_f$, we now solve the truncated rotational model of Equation (\ref{eq:3by3}).  For the estimate of viscous friction at the CMB, we set $Re[K_{cmb}]=0$ and use $Im[K_{cmb}]$ as given by Equation (\ref{eq:kcmbt2}) with the latest estimate of $(K/C)$ equal to $1.41\pm0.34 \times 10^{-8} \ \mathrm{days}^{-1}$, as inferred from LLR \cite[][]{williams15}.  With this choice, and in the absence of tidal deformation ($S_{ij} \rightarrow 0$), viscous coupling at the CMB by itself results in a phase lead of   the symmetry axis of the mantle with respect to the Cassini plane of approximately $\phi_p = 0.123$ arcsec ($\sim46$\% of the observed 0.27 arcsec). This is largely insensitive to the viscosity of the LVZ or the CMB radius.  (This is perhaps counter intuitive given that $K_{cmb}$ depends on $\bar{A}_f$ in Equation (\ref{eq:kcmbt2}). But recall that for each choice of core radius, the densities of the fluid core and mantle/LVZ are adjusted to match $\bar{\rho}$ and $I_{sm}$.  Thus $\bar{A}_f$ changes little between different choices of $r_f$).  Tidal deformation must account for the remaining 0.147 arcsec ($\sim54$\%) to match the observed phase lead of 0.27 arcsec.  

\begin{figure}
\begin{center}
    \includegraphics[height=8cm]{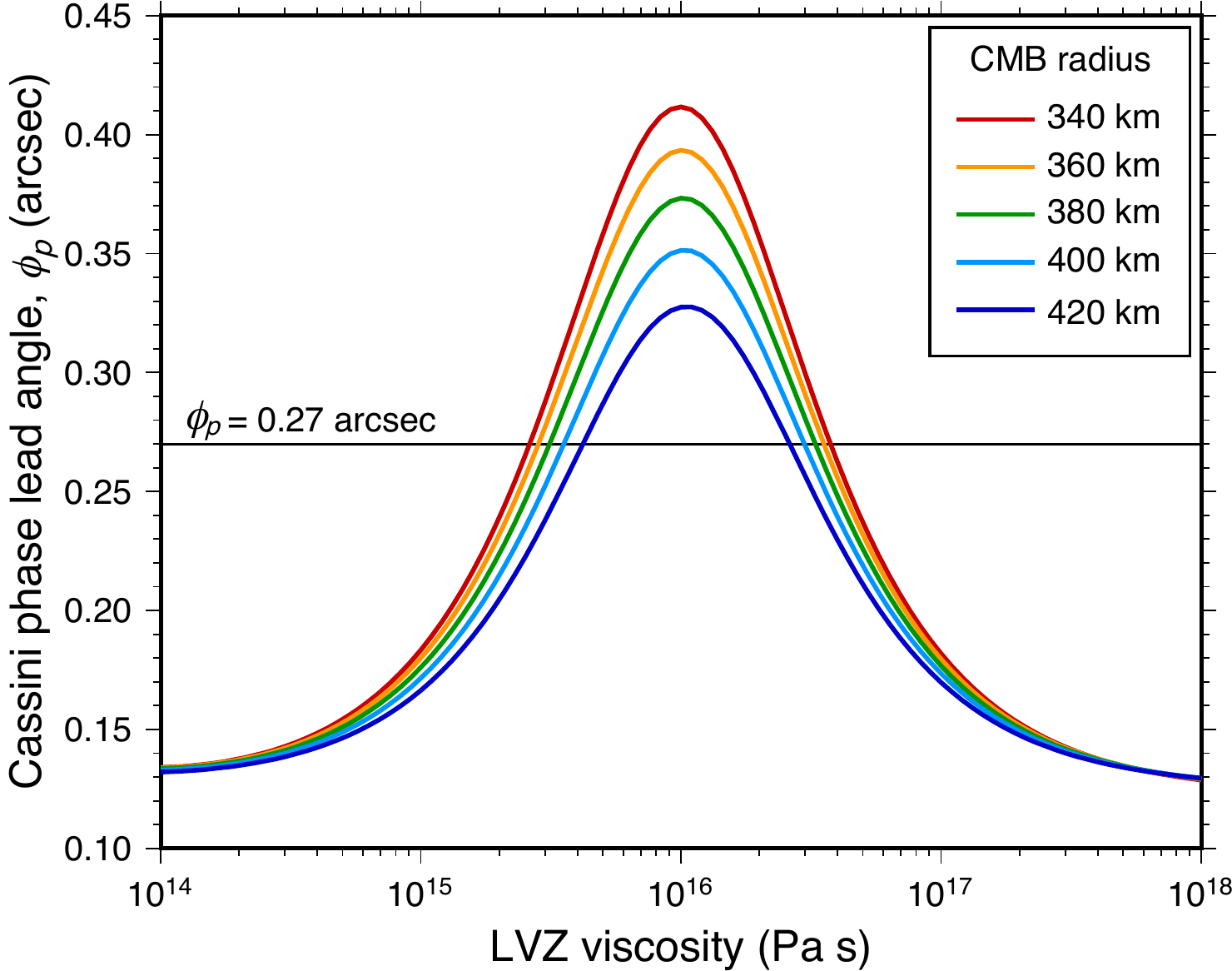}  
    \caption{\label{fig:lead_nosic} The phase lead angle $\phi_p$ of the figure axis of the mantle ahead of the Cassini plane  as a function of LVZ viscosity for different choices of CMB radius.  The LVZ outer radius is 550 km and there is no inner core. }
\end{center}
\end{figure}

Figure \ref{fig:lead_nosic} shows how $\phi_p$ varies as a function of $\eta_{lvz}$ and $r_f$.  $\phi_p$ is largest when tidal dissipation within the lunar mantle is largest, so it tracks the imaginary part of $S_{11}$.  The maximum energy dissipation, and thus the largest phase lead, occurs for a range of LVZ viscosities centred around approximately $10^{16}$ Pa s. For LVZ viscosities of approximately $3-4 \times 10^{16}$ Pa s, the combination of tidal dissipation and viscous friction at the CMB can match the observed phase lead of 0.27 arcsec.

\cite{harada16} have shown that a LVZ layer with an outer radius of 550 km and the rheology of a Maxwell solid with a viscosity of approximately $3 \times 10^{16}$ Pa s can explain the monthly value of $Q=38\pm4$.  Since this $Q$ value is itself inferred from LLR observations in order to fit $\phi_p$, it should then come as no surprise that, as we show in Figure \ref{fig:lead_nosic}, our rotational model should retrieve the observed $\phi_p$ with a LVZ layer of similar characteristics.  Nevertheless, that we can retrieve this result serves as a good test of the consistency of our rotational model. It shows that it captures the essential dissipation ingredients of the Cassini state.  

Because we focus on the Cassini state, our rotational model is simpler than the ones used in LLR studies.  This allows us to characterized more straightforwardly the relative importance of viscous friction at the CMB and tidal dissipation to $\phi_p$, and how these may vary as a function of lunar model.  A prediction for $\phi_p$ can be constructed from the truncated rotational model of Equation (\ref{eq:3by3}).  To a very good approximation, the curves shown in Figure  \ref{fig:lead_nosic} are determined by 

\begin{equation}
\phi_p =  \left( \frac{1}{\delta \omega - \beta \Phi_2} \right) \bigg[ Im[S_{11}] \left( \Phi_1^t + \Phi_2^t Re[\tilde{p}] \right) - \frac{\bar{A}_f}{\bar{A}} \left( \frac{ \delta \omega }{e_f - \delta \omega } \right)^2 \, Im[K_{cmb}] \, Re[\tilde{p}]  \bigg] \, .\label{eq:varphip_pred1}
\end{equation}  
The first and second terms in the square bracket on the right-hand side represent the contributions from tidal dissipation and viscous friction at the CMB, respectively.  It is useful to rewrite this expression in terms of $k_2/Q$ and $K/C$, as these are the parameters that are reported in LLR studies \cite[][]{williams01,williams14,williams15},

\begin{equation}
\phi_p = \left( \frac{1}{\delta \omega - \beta \Phi_2} \right) \bigg[ \left( \frac{k_2}{Q} \right) \frac{R^5\,  \Omega_o^2 \, \Phi^t }{3 G \bar{A}}  + \left(\frac{K}{C} \right)\left(\frac{ \delta \omega }{e_f - \delta \omega } \right)^2  \frac{\sin(\theta_p)}{\Omega_o}  \bigg] \, , \label{eq:varphip_pred2}
\end{equation}
where $\Phi^t$ is given by Equation (\ref{eq:Phit}).  To our knowledge, such a direct expression connecting $\phi_p$ with $k_2/Q$ and $K/C$ has not been presented before.  

We can readily verify that the prediction of Equation (\ref{eq:varphip_pred2}) is correct.  Using $I=5.145^\circ$ and $\theta_p=1.543^\circ$, then independent of the choice of interior model, $\Phi_2=1.4646$ and $\Phi^t=0.5523$.  $\beta$ and $e_f$ depend on the interior model; for a lunar core of radius $r_f=380$ km, and with our choice of crustal density and thickness, $\beta = 6.314\times 10^{-4}$ and $e_f = 2.153 \times 10^{-4}$.  Using the latest estimates of $k_2/Q = (6.4\pm 1.5)\times 10^{-4}$ and $K/C=(1.41\pm0.34)\times 10^{-8}$ day$^{-1}$ \cite[][]{williams15}, the tidal dissipation and CMB friction estimated from Equation (\ref{eq:varphip_pred2}) contribute respectively $(0.151\pm0.035)$ and $(0.123\pm0.030)$ arcsec to the lead angle $\phi_p$.  Adding the central value for each of these contribution gives $\phi_p =0.274$ arcsec, in general agreement with LLR observations.

Note that in our linear rotational model, only the perturbations at a monthly period affect the resulting $\phi_p$.  In reality, perturbations at other frequencies can interact together non-linearly to yield a contribution to $\phi_p$ at a monthly period.   These non-linear interactions are taken into account in LLR studies, and small difference are expected with our prediction of $\phi_p$.

The appeal of Equation (\ref{eq:varphip_pred2}) is that one can test whether a specific combination of $k_2/Q$ and $K/C$ is consistent with $\phi_p=0.27$ arcsec without having to perform an analysis of LLR observations.  Obviously, $\phi_p$ itself must be determined by LLR.  Likewise, the relative contributions of tidal dissipation and CMB friction can only be recovered by LLR observations of the lunar librations at different frequencies.  But if a theoretical model of $k_2/Q$ is proposed, the required amplitude of $K/C$ can be inferred from Equation (\ref{eq:varphip_pred2}), and vice-versa.

\section{Results II: Viscoelastic relaxation within the inner core}

We now investigate the role of the inner core in the dissipation associated with the Cassini state of the Moon.  If a solid inner core is present at the centre of the Moon, its rotation vector is expected to be misaligned relative to both the mantle and fluid core rotation vectors (DW16, SD18).  Viscous friction from the differential velocity at the ICB adds a new source of dissipation.  The ratio of the ICB to CMB friction torque is 

\begin{equation}
\frac{\tilde{\Gamma}_{icb}}{\tilde{\Gamma}_{cmb}} = \frac{K_{icb}}{K_{cmb}} \left( \frac{r_s}{r_f} \right)^5 \frac{\tilde{m}_s - \tilde{m}_f}{\tilde{m}_f} \, .
\end{equation}

Although for an inner core with a radius half of that of the outer core the factor $(r_s/r_f)^5$ equals $1/32$, the viscous torque at the ICB may nevertheless contribute non-negligibly if the tilt of the rotation vector of the inner core is much larger than that of the fluid core, and hence if $\| (\tilde{m}_s-\tilde{m}_f)/\tilde{m}_f \| > 1$.  Moreover, for a turbulent friction model the coupling parameters $K_{icb}$ and $K_{cmb}$ are themselves proportional to the differential velocities, and the torque ratio would then involve $\| (\tilde{m}_s-\tilde{m}_f)/\tilde{m}_f\|^2$.  The tilt angle of the rotation vector of the inner core is unknown.  However, because the free inner core nutation (FICN) frequency of the Moon is close to the frequency of the 18.6 yr precession, the offset between the inner core and mantle rotation vectors can be large (DW16, SD18).  Hence, it is possible that viscous friction at the ICB contributes significantly to the viscous dissipation inferred by LLR. In this case, LLR would detect an effective friction from the combined effects at both the CMB and the ICB.   

Despite the potential importance of friction at the ICB, our focus here is instead on the role of viscoelastic deformation within the inner core.  We consider  an end member scenario where friction at both the ICB and CMB contribute negligibly to $\phi_p$.  That is, we set $K_{cmb}=K_{icb}=0$ in our rotational model of Equation (\ref{eq:5by5}).  For all interior models in this section, the LVZ radius and viscosity are set at 550 km and $3 \times 10^{16}$ Pa s. The viscosities of the mantle and crust remain fixed at $1 \times 10^{20}$ Pa s.  

\subsection{Viscoelastic deformation}

\begin{figure}
\begin{center}
    \includegraphics[height=12cm]{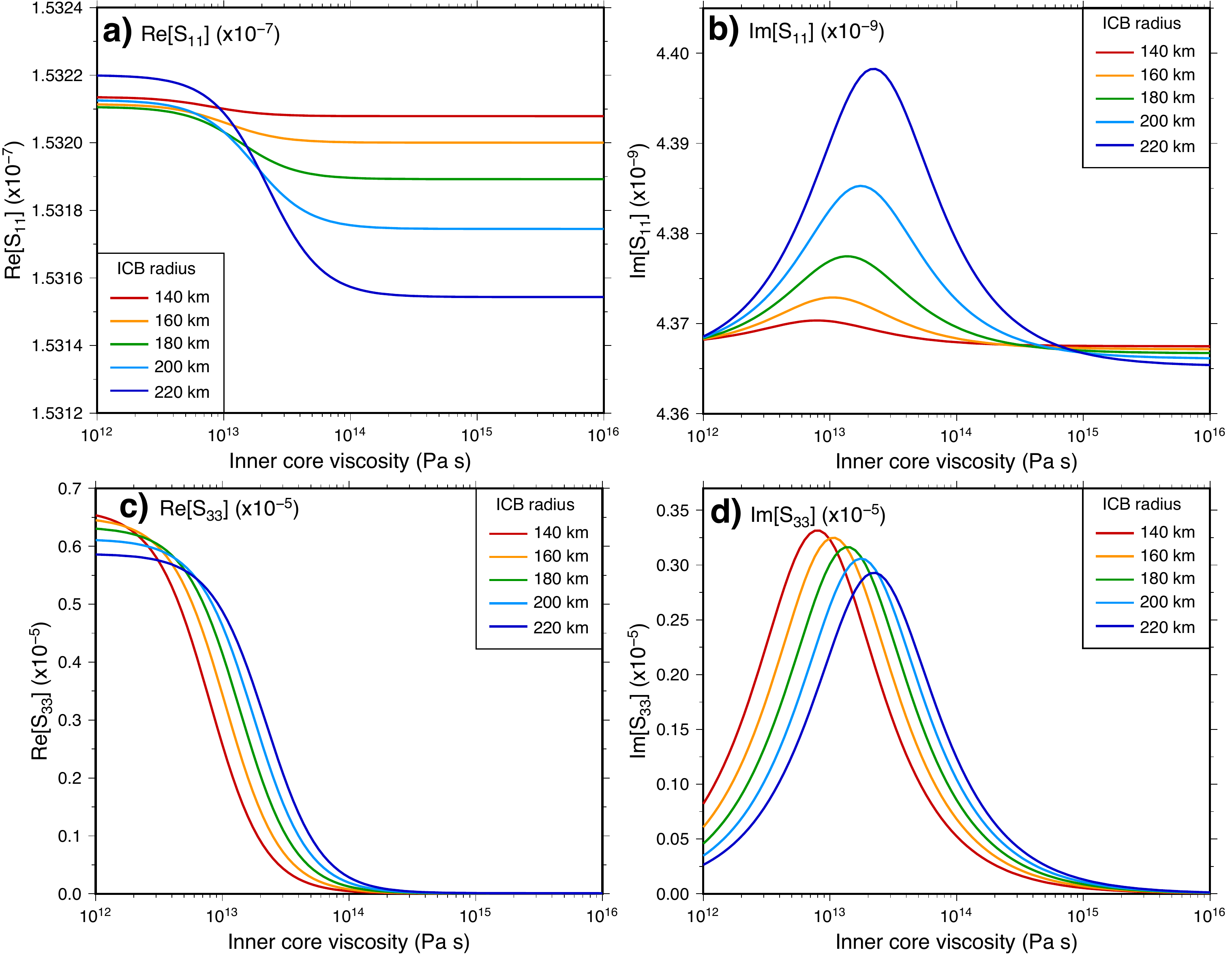} 
    \caption{\label{fig:S11S33} a) Real and b) imaginary parts of the compliance $S_{11}$, and c) real and d) imaginary parts of the compliance $S_{33}$ as a function of inner core viscosity and for different choices of inner core radius.  In each case, the CMB radius is 380 km, the LVZ outer radius is 550 km and the LVZ viscosity is $3 \times 10^{16}$ Pa s.}
\end{center}
\end{figure}

We investigate first how viscoelastic relaxation within the inner core affects the compliance $S_{11}$. We set the CMB radius to 380 km and we build interior models for a range of inner core radii between 140 and 220 km and inner core viscosities ($\eta_s$) between $10^{12}$ and $10^{16}$ Pa s.  Figures \ref{fig:S11S33}a,b show how $S_{11}$ changes as a function of $\eta_s$ for different $r_s$.  The general behaviour of both the real and imaginary parts of $S_{11}$ is similar to that produced by viscoelastic deformation within the LVZ shown in Figures \ref{fig:S11_nosic}a,b.  Energy dissipation is maximum for $\eta_s \approx 1-2 \times 10^{13}$ Pa s and, unsurprisingly, is largest for the model with the largest inner core radius of 220 km.  But note how the change in $S_{11}$ amplitude remains very small compared to its base value.  For $r_s=220$ km, the maximum changes in $Re[S_{11}]$ and $Im[S_{11}]$ in Figures \ref{fig:S11S33}a,b are approximately $6 \times 10^{-11}$ and $3 \times 10^{-11}$, respectively.  This is more than 200 times smaller than the part of $S_{11}$ resulting from deformation within the LVZ.

This demonstrates that, mainly because of its small volume, the global deformation of the Moon is largely insensitive to the presence of an inner core.  This result is important because it shows that the presence of an inner core does not contribute significantly, even if its viscosity is small, to the global tidal dissipation estimated through the parameter $k_2/Q$.  Hence, if viscoelastic deformation within the inner core contributes to the dissipation associated with the Cassini state of the Moon, it is by taking a share of the part which is currently assumed to be from friction at the CMB; the part associated with the term $K/C$ in Equation (\ref{eq:varphip_pred2}). 

How the inner core deforms in response to a forcing that is applied solely to the inner core is captured by the compliance $S_{33}$. Figures \ref{fig:S11S33}c,d show how $S_{33}$ changes as a function of $\eta_s$ and $r_s$.  The behaviour of $S_{33}$ is again characteristic of a Maxwell solid.  The inner core behaves like an elastic solid when $\eta_s > 10^{15}$ Pa s and like a fluid when $\eta_s < 10^{12}$ Pa s.  Energy dissipation is maximum for $\eta_s \approx 10^{13}$ Pa s.

The inner core viscosity at which the fluid-to-elastic transition occurs (and at which energy dissipation is maximum) is determined by the characteristic viscous relaxation time $\tau_s$ of the inner core, often referred to as the Maxwell time.  An estimate of $\tau_s$ is given by \cite{buffett97},

\begin{equation}
    \tau_s \approx \frac{\chi \, \eta_s}{(\rho_s-\rho_f) \, g_s \, r_s} \, , \label{eq:taus}
\end{equation}
where $g_s$ is the gravitational acceleration at the ICB and $\chi$ is a numerical constant of order unity.  When $\tau_s$ is much longer than the period of the forcing (one month, for the Cassini state of interest here), the inner core does not have sufficient time to relax viscously in response to the imposed forcing and behaves like an elastic solid. Conversely, when $\tau_s$ is much shorter than the forcing period, the inner core has time to fully relax in response to the imposed forcing and behaves essentially like a fluid.  Dissipation is maximized when $\tau_s$ is equal to the period of the forcing, and the viscosity of the inner core which is required for $\tau_s$ given by Equation (\ref{eq:taus}) to be equal to approximately one month is of the order of $10^{13}$ Pa s.   Equation (\ref{eq:taus}) also shows that a larger inner core requires a larger  viscosity to yield the same $\tau_s$, and this explains the shift in $\eta_s$ at which the maximal value of $Im[S_{33}]$ for different $r_s$ occurs in Figure \ref{fig:S11S33}d.

\subsection{The Phase Lead of the Cassini State}

We now investigate how viscoelastic deformation within the inner core may contribute to the phase lead angle $\phi_p$.  Since $k_2/Q$ is largely unaffected by the presence of an inner core, tidal dissipation within the mantle must remain an important contribution to $\phi_p$.  As shown in the previous section, viscoelastic deformation concentrated within a LVZ layer of outer radius equal to 550 km and a viscosity of $3 \times 10^{16}$ Pa s can account for approximately 0.15 arcsec of the total phase lead. We seek to find the requirements on inner core size and viscosity so that viscoelastic deformation within the inner core can explain a portion or the totality of the remaining part of the observed phase lead. 

\begin{figure}
\begin{center}
    \includegraphics[height=8cm]{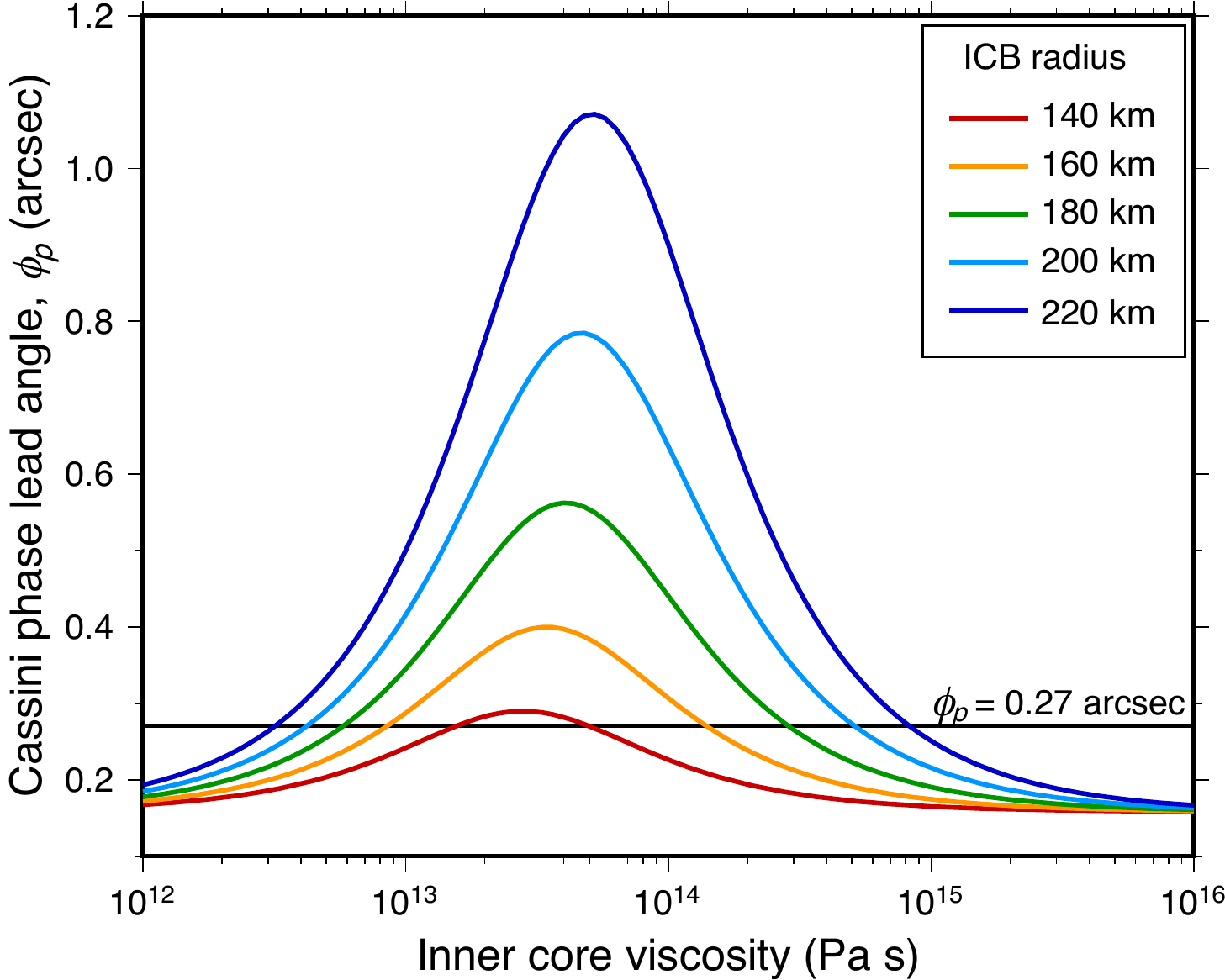}  
    \caption{\label{fig:imp} The phase lead angle $\phi_p$ of the figure axis of the mantle ahead of the Cassini plane as a function inner core viscosity and for different choices of inner core radii. }
\end{center}
\end{figure}

We first show a set of results for a fixed outer core radius of $r_f=380$ km.  For each combination of $\eta_s$ and $r_s$, we calculate the full set of compliances and solve the 5-by-5 rotational model of Equation (\ref{eq:5by5}).   Figure \ref{fig:imp} shows how $\phi_p$ changes as a function of $\eta_s$ for different $r_s$.  The larger the inner core is, the greater the contribution to $\phi_p$ is from inner core relaxation.  For an inner core larger than 140 km, the observed $\phi_p$ can be matched provided $\eta_s$ is in the range of approximately $3 \times 10^{12}$ to $10^{15}$ Pa s.  More specifically, for each choice of $r_s > 140$ km, two different values of $\eta_s$ permit a match of the observed $\phi_p$, one on each side of the dissipation peak.  Figure \ref{fig:imp} further shows that if $\eta_s$ is such that maximum viscous dissipation occurs within the inner core (in other words,  $\eta_s$ such that the Maxwell time is close to one month), the phase lead angle can exceed significantly the observed $\phi_p =0.27$ arcsec.

The manner in which the inner core affects $\phi_p$ is not through its delayed response to the tidal forcing from Earth; this is included in the global compliance $S_{11}$, and as we have shown above, the relative contribution of the inner core is small.   Instead, it is by an exchange of angular momentum between the inner core and the mantle, operating through the gravitational torque between their misaligned figures.  In the Cassini state, the tilt angle of the figure axis of the inner core is misaligned with respect to that of the mantle (DW16, SD18).  Torques acting on the inner core are responsible for this tilt, and the largest of these is the gravitational torque due to the mantle. (The total gravitational torque acting on the inner core also includes a contribution from the fluid core, but the dominant contribution is from the mantle).  A viscoelastic inner core deforms to realign its figure with the gravitational potential imposed by the mantle, though with a time delay.  Hence, the rotation vector of the  inner core acquires a component out of the Cassini plane; specifically, a positive imaginary component.  To conserve angular momentum, the spin-symmetry axis of the mantle figure acquires a tilt in the reverse direction, a negative imaginary component, contributing to its lead angle $\phi_p$.  

This can be understood on the basis of the angular momentum balance for the whole Moon expressed by Equation (\ref{eq:am1}).  The largest contribution to the angular momentum from the inner core is contained in its rotation vector, in the term $(1+\omega) (\bar{A_s}/\bar{A} )\,\tilde{m}_s = - \delta \omega (\bar{A_s}/\bar{A})\,\tilde{m}_s$.  Keeping only the out-of-plane contribution from this term for the angular momentum of the whole core, along with tidal dissipation of the whole Moon, the prediction for $\phi_p$ is 

\begin{equation}
\phi_p =   \left( \frac{1}{\delta \omega - \beta \Phi_2} \right) \bigg[ Im[S_{11}] \left( \Phi_1^t + \Phi_2^t Re[\tilde{p}] \right) + \frac{\bar{A}_s}{\bar{A}} \delta \omega \, Im[\tilde{m}_s] \bigg] \, .\label{eq:phip_pred_sic}
\end{equation}  
This prediction is a very good approximation to the solutions shown in Figure \ref{fig:imp}, to the point that on the scale of the figure they are indistinguishable.  

\begin{figure}
\begin{center}
    \includegraphics[height=6cm]{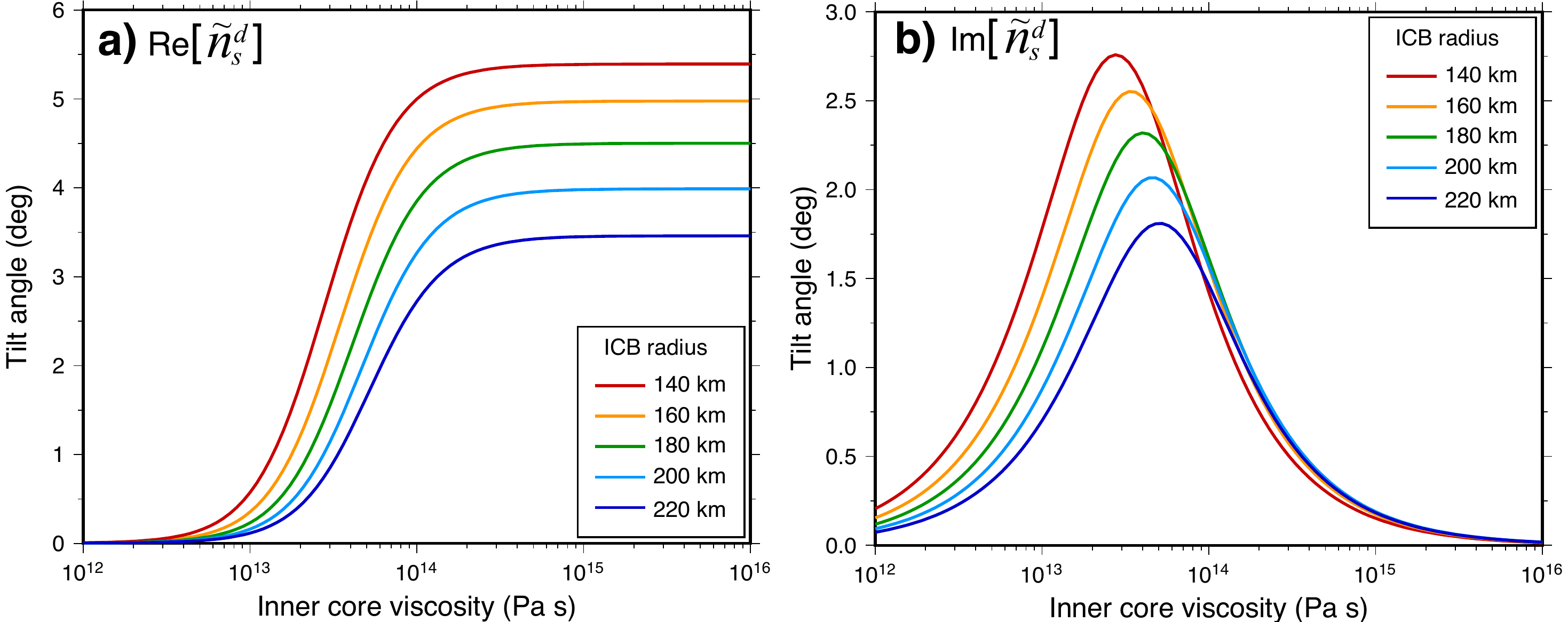} 
    \caption{\label{fig:impnsd} a) Real and b) imaginary parts of the tilt angle of the figure axis of the deformed inner core $\tilde{n}_s^d$ (in degrees) as a function of inner core viscosity and for different inner core radii.}
    \end{center}
\end{figure}

Furthermore, if the inner core and mantle interact through gravitational coupling, we expect a correlation between $\phi_p$ and the tilt angle of the instantaneous figure axis of the deformed inner core.  The latter, which we denote by $\tilde{n}_s^d$, involves $\tilde{n}_s$ (the figure axis of the undeformed inner core) and $\tilde{c}_s$ (the off-diagonal elements of the moment of inertia tensor) and is given by \cite[e.g.][]{dumberry09}

\begin{equation}
\tilde{n}_s^d = \tilde{n}_s +  \frac{\tilde{c}_s}{A_s e_s} \approx \tilde{n}_s \bigg( 1 - \frac{S_{33} \alpha_3 \alpha_g}{e_s} \bigg) \, ,
\end{equation}
where in the approximation on the right-hand side we have assumed that the gravitational potential $\tilde{\phi_s^g}$ (Equation \ref{eq:phiscg}) dominates all contributions to $\tilde{c}_s$. Figure \ref{fig:impnsd} shows how the real and imaginary parts of $\tilde{n}_s^d$ vary as a function of $\eta_s$ and $r_s$.  In the elastic limit ($\eta_s > 10^{16}$ Pa s), $Im[\tilde{n}_s^d ] \rightarrow 0$, and the figure axis of the inner core lies in the Cassini plane.  Its tilt angle depends on $r_s$  through the way in which the latter influences the FICN frequency (DW16, SD18).  For the case illustrated in Figure \ref{fig:impnsd}, a decreasing inner core radius brings the FICN frequency closer to the forcing frequency (i.e. the precession frequency $\Omega_p$), and a larger inner core tilt results from resonant amplification.  In the fluid limit ($\eta_s < 10^{12}$ Pa s), the figure of the deformed inner core is aligned with the mantle.  For $\eta_s$ in the range $10^{13}$-$10^{15}$ Pa s, at the transition between the fluid and elastic regimes, the real and imaginary parts of $\tilde{n}_s^d$ are of similar magnitudes; the figure axis of the deformed inner core no longer lies in the Cassini plane, but lags behind it significantly ($Im[\tilde{n}_s^d] > 0$).  The $\eta_s$ values at which the peaks of $Im[\tilde{n}_s^d]$ are located (Figure \ref{fig:impnsd}) are correlated with the location of the peaks of $\phi_p$ (Figure \ref{fig:imp}), further corroborating that it is through gravitational coupling that viscoelastic inner core deformation contributes to $\phi_p$. Note that $Im[\tilde{n}_s^d]$ is of the order of a few degrees, much larger than  the lead angle of the mantle that it produces (of the order of a fraction of an arcsec), but this is because of the large contrast in their moments of inertia ($\bar{A}_s \ll \bar{A}$).

\begin{figure}
\begin{center}
   \includegraphics[height=6.5cm]{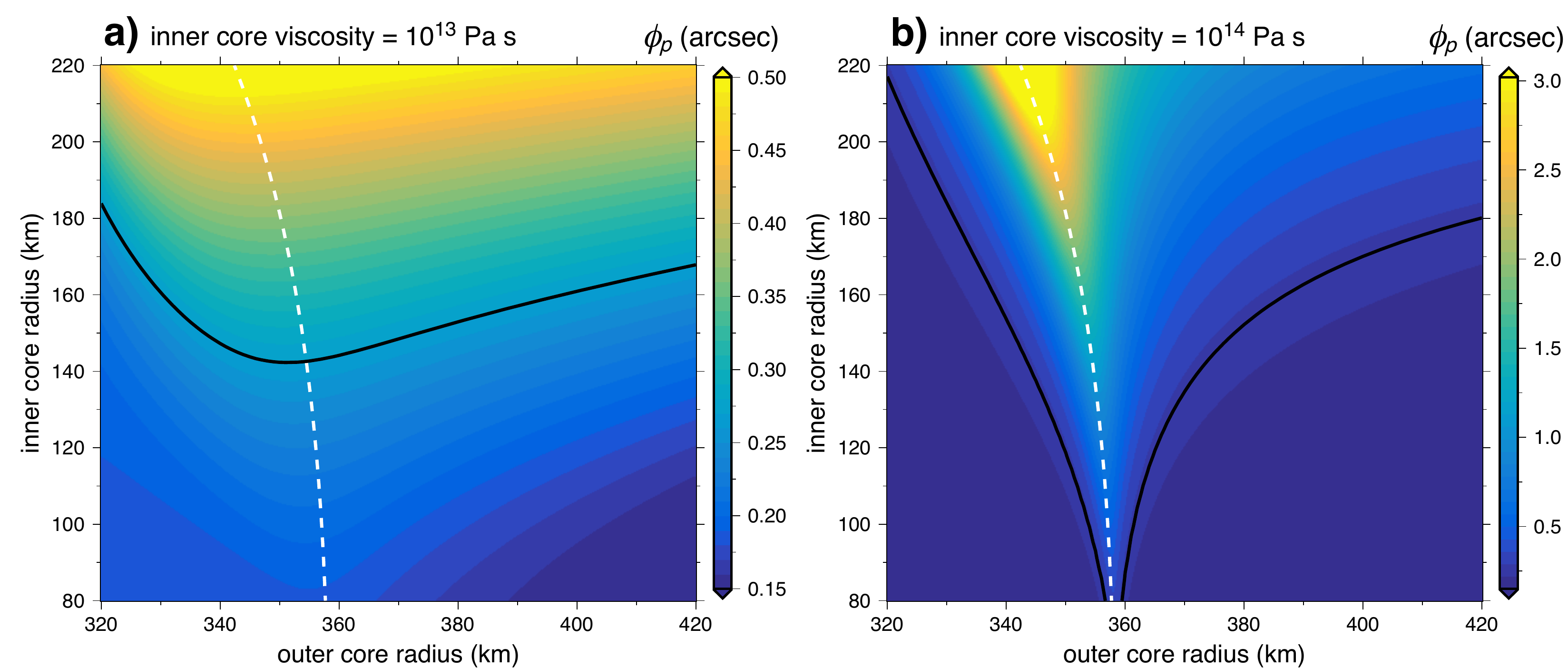} 
        \caption{\label{fig:rsrfphip} 
The phase lead angle $\phi_p$ of the figure axis of the mantle ahead of the Cassini plane (colour contours, in arcsec) as a function of inner core and outer core radii, for an inner core viscosity $\eta_s$  of a) $10^{13}$ Pa s and b) $10^{14}$ Pa s.  The black contour line corresponds to the observed phase lead of 0.27 arcsec.  Lunar models for which the FICN frequency $\Omega_{ficn}$ is equal to the precession frequency $\Omega_p = 2\pi/18.6$ yr$^{-1}$ follow the white dashed contour line.  In b), the colour scale is saturated at 3 arcsec.}
\end{center}
\end{figure}

The results shown in Figures \ref{fig:imp} and \ref{fig:impnsd} pertain to a fluid core radius of $r_f =380$ km.  Figure \ref{fig:rsrfphip} shows how $\phi_p$ varies as a function of $r_f$ and $r_s$ for two different choices of $\eta_s$: $10^{13}$ and $10^{14}$ Pa s.  These contour plots further confirm that the larger the inner core radius is, the larger the phase lead angle $\phi_p$ it can generate. But they also highlight that inner core size is not the only factor.  The choice of $r_f$ also influence the resulting $\phi_p$.  This is because the magnitude of $\tilde{n}_s^d$ depends on how close the FICN frequency ($\Omega_{ficn}$) is to $\Omega_p$.  A good approximation of $\Omega_{ficn}$, as seen in the inertial frame, is (DW16)

\begin{equation}
    \Omega_{ficn} \approx \Omega_o e_s (-\alpha_1 + \alpha_3 \alpha_g + \alpha_3 \Phi_2) \, . \label{eq:ficn}
\end{equation}
The FICN mode is retrograde, but note that we have defined the retrograde direction to be positive here, the same convention that we used when we defined $\Omega_p$, and hence $\Omega_{ficn}$ can be directly compared to $\Omega_{p}$.  The term involving $\alpha_g$ captures gravitational coupling between the inner core and the rest of the Moon and it is the largest of the three terms on the right-hand side.  This term involves $\alpha_3$, a measure of the density contrast at the ICB.  Thus the density of the fluid core, which depends on the choice of $r_f$ in our model, directly influences the FICN frequency.  

The combinations of $r_s$ and $r_f$ that yield $\Omega_{ficn}$ equal to $\Omega_p = 2\pi/18.6$ yr$^{-1}$ are indicated on Figure \ref{fig:rsrfphip} by a white dashed contour line. The largest $\phi_p$ values are achieved when the $\Omega_{ficn}$ is closest to $\Omega_p$.  This is because, by resonant amplification, the in-phase tilt angle of the inner core is largest when $\Omega_{ficn}$ is closest to $\Omega_p$.  The larger the in-phase tilt angle is, the larger its out-of-plane component can be as a result of viscoelastic deformation (see Figure \ref{fig:impnsd}).  Hence, the contribution to $\phi_p$ from inner core relaxation depends not only on the size and viscosity of the inner core, but also on the proximity of $\Omega_{ficn}$ to $\Omega_p$.    

Figure \ref{fig:rsrfphip} shows that, provided the viscosity of the inner core is in the range $10^{13}-10^{14}$ Pa s, viscoelastic deformation may contribute to a part of the observed $\phi_p$.  Indeed, it is possible to explain the entire observed phase lead angle of $\phi_p=0.27$ arcsec without a contribution from viscous friction at the CMB and ICB.  Furthermore, as pointed out when describing Figure \ref{fig:imp}, specific combinations of $r_s$, $r_f$ and $\eta_s$ lead to a prediction of $\phi_p$ that far exceeds its observed value; such combinations would be incompatible with observations.  For instance, if the inner core viscosity is approximately $10^{13}$ Pa s, then its radius cannot be larger than approximately 180 km.  This upper bound is of course reduced the greater the fraction of $\phi_p$ which is due to viscous friction at the CMB and ICB, which is set to zero in these calculations.  Conversely, if the inner core is large, our results indicate that its viscosity cannot be too low, and $\Omega_{ficn}$ cannot be too close to $\Omega_p$, as otherwise $\phi_p$ would exceed 0.27 arcsec.

As the inner core viscosity is increased beyond $10^{14}$ Pa s, the contour levels of $\phi_p$ get more concentrated closer to the FICN resonance.  This is because as $\eta_s$ increases, the ratio of the real to imaginary components of the inner core tilt also increases. Hence, a larger in-phase  tilt angle is required to produce the same out-of-phase tilt amplitude.  Because of the reduced dissipation within the inner core, the amplitude of the inner core tilt near the resonance gets very large, in excess of $30^\circ$. Since our model is built under the assumption of small angles, our results are not very accurate close to the FICN resonance when $\eta_s \ge 10^{14}$ Pa s.   In fact, this is also true for inner core radii smaller than 100 km for the case of $\eta_s=10^{14}$ Pa s shown in Figure \ref{fig:rsrfphip}b.  In our linear model, without any dissipation, the tilt angle of the inner core diverges to $\pm \infty$ when $\Omega_{ficn}$ is equal to $\Omega_p$ (DW16).  However, as shown in SD18, in a model valid for large angles, though still in the absence of dissipation, the in-phase inner core tilt angle does not diverge to infinity near the FICN resonance, but is instead bound to be between $-33^\circ$ and $17^\circ$.  This places an upper bound for the largest in-phase inner core tilt.  With increasing $\eta_s$, even if the in-phase inner core tilt amplitude is close to one of these extrema, the out-of-phase component would at some point be too small to generate a significant contribution to $\phi_p$.  Hence, there is an upper bound value of $\eta_s$ beyond which viscoelastic inner core deformation no longer plays a significant role in the observed $\phi_p$.  It is difficult to give a precise measure of this upper bound based on our linear model.  But from inspection of our results, we estimate that inner core deformation contributes less than 0.01 arcsec to the observed $\phi_p$ when $\eta_s >10^{15}$ Pa s for $r_s=100$ km in radius, and when $\eta_s >10^{17}$ Pa s for $r_s=200$ km. 

Lastly, the $\phi_p$ contour maps of Figure \ref{fig:rsrfphip} are tied to the choices we have made for the density and thickness of the crust. These influence the densities of the mantle, LVZ, and core that can match $I_{sm}$ and $\bar{\rho}$ (see Equation \ref{eq:rhoIsm}), and in turn, this affects the frequency of the FICN for a given combination of $r_s$ and $r_f$.  With different assumptions about the crust, the location of the FICN resonance would be shifted on Figure \ref{fig:rsrfphip} and so would the $\phi_p$ contours.  Our general conclusions remain unaltered, but one should be careful in extracting specific values of $\phi_p$ as a function of $r_s$ and $r_f$ from Figure \ref{fig:rsrfphip}. 

\section{Discussion and Conclusion}

We have presented in this work an extension of the model of the Cassini state of the Moon developed in DW16 and SD18 to include dissipation.  Specifically, the two forms of dissipation that are included in our model are viscous coupling at the boundaries of the fluid core and viscoelastic deformation within the solid regions of the Moon.  Our model allows us to investigate how these contributions can account for the dissipation observed by LLR. In particular, how this dissipation manifests itself through a phase lead angle of $\phi_p=0.27$ arcsec in the spin-symmetry axis of the Moon relative to the plane of an exact Cassini state.  We have focused our attention on the role of viscoelastic deformation occurring within a LVZ layer at the bottom of the mantle and within the solid inner core.

In accordance with the studies of \cite{harada14,harada16}, we have shown that viscoelastic deformation concentrated within a LVZ at the base of the mantle, with an outer radius equal to 550 km and a viscosity of approximately $3 \times 10^{16}$ Pa s, is consistent with the monthly $Q$ value inferred by LLR.  The global tidal dissipation from this effect accounts for approximately 0.15 arcsec of the observed $\phi_p$.  Furthermore, we have shown that the inner core plays a negligible role in the global tidal dissipation.  

Our results also demonstrate that viscoelastic inner core deformation can contribute significantly to the observed $\phi_p$ through gravitational coupling with the mantle.  The precise value of $\phi_p$ depends on the inner core and fluid core radii and also on the inner core viscosity.  The larger the inner core is, and the closer the FICN frequency is to being in resonance with the precession frequency, the larger is the contribution from inner core relaxation to $\phi_p$.  Maximal contribution to $\phi_p$ results with an inner core viscosity of the order of $10^{13}$ to $10^{14}$ Pa s, but inner core relaxation may contribute non-negligibly to $\phi_p$ for a viscosity as large as $10^{16}$ Pa s.  

Whether the Moon has an inner core remains uncertain.  If an inner core is present, its viscosity is not known.  Hence, whether the mechanism that we have investigated can realistically contribute to dissipation in the Moon remains speculative.  However we can ask how a viscosity of the order of $10^{13}$ to $10^{15}$ Pa s compares with estimates that have been suggested for the viscosity of the Earth's inner core.   A few estimates have been obtained from geodynamic modelling of the rotational dynamics of the Earth's inner core.  Considerations of its differential rotation \cite[][]{buffett97}, its possible axial oscillations \cite[][]{dumberry10b,davies14} and its contribution to the observed nutations \cite[][]{greff00,koot11} suggest an inner core viscosity in the range of $10^{15}-10^{17}$ Pa s.  Laboratory experiments on the strength of iron suggest an inner core viscosity in a similar range of $10^{15}-10^{18}$ Pa s \cite[][]{gleason13}. The recent detection of seismic $J$-waves \cite[][]{tkalcic18} is further suggestive of an inner core viscosity within this range.   Relative to the Earth's inner core, the inner core of the Moon is at a lower pressure and temperature, so we must be careful in mapping estimates obtained for the Earth.  Nevertheless, with this caveat in mind, a viscosity close to or even smaller than  $10^{15}$ Pa s is not inconceivable, and hence viscoelastic inner core relaxation may indeed play a role in the dissipation observed in the lunar rotation dynamics.

We have focused our attention on viscoelastic deformation, but another process that may act to realign the shape of the inner core with that of the mantle is melting and crystallization.  Within the fluid core, hydrostatic equilibrium implies that surfaces of constant gravitational potential, density and pressure are all aligned. Surfaces of constant temperature also follow this alignment because temperature is linked to pressure and density by an equation of state.  Since the ICB marks the solid-liquid phase transition, at the mean radius of the inner core, the temperature should coincide with the liquidus (melting temperature).  However, a tilted ellipsoidal inner core has its ICB misaligned relative to the liquidus (e.g. see figure 8 of SD18).  The parts of the inner core surface that are at a higher temperature undergo melting, while the parts that are at a lower temperature experience crystal growth.  This melt and growth process operates to realign the shape of the ICB toward an alignment with the mantle and hence, may contribute to an effective viscoelastic deformation of the inner core shape.  

It is important to stress that we do not suggest that viscous friction at the CMB and ICB does not play an important role in the observed dissipation.  Although we have considered an end-member scenario in which these effects are absent, this was simply to isolate the possible role of viscoelastic inner core deformation.  In all likelihood, viscous friction at the solid boundaries of the fluid core contribute to dissipation.  But the question is, by how much?  To fit LLR observations, so far the only rotational models of the Moon that have been considered do not have an inner core.  Viscous friction at the CMB must then account, by itself, for the part of the dissipation not accounted for by tidal dissipation.  The magnitude of the CMB friction torque that is retrieved is broadly in line with the expected magnitude of a viscous torque in a turbulent regime \cite[][]{yoder81,williams01}, lending further support to this interpretation.  But a direct prediction of the magnitude of the turbulent friction torque at the CMB is not possible because it involves a numerical factor which depends on surface roughness, which is unknown. Moreover, with an inner core present, viscous friction takes place at both the CMB and ICB.  Depending on the orientation of the inner core in the Cassini state, the friction torques at the CMB and ICB may add up together or partly cancel one another.   Our point is that sufficient uncertainty exists in estimating the friction torque at the CMB and ICB that there is room for viscoelastic inner core deformation to contribute to the observed dissipation.

We have focused our efforts here on the dissipation associated with the Cassini state, but viscoelastic relaxation within the inner core would also participate in the dissipation of the forced and free lunar librations.  According to the Maxwell time relation of Equation (\ref{eq:taus}), the longer the libration period, the larger is the inner core viscosity at which maximum dissipation occurs.  This implies that viscoelastic deformation within the inner core could play a proportionally more important role for librations at longer periods. For instance, in the case of the longitudinal libration with a period of 6 years, the curves for the imaginary part of $S_{33}$ on Figure \ref{fig:S11S33}d are shifted to higher viscosities: the dissipation peak would be moved from $\sim10^{13}$ to $\sim10^{15}$ Pa s.  For an inner core viscosity of, say, $10^{16}$ Pa s, inner core relaxation may play an important role in the energy dissipation associated with the 6 year libration, even though its contribution to the lead angle $\phi_p$ of the Cassini state (27.212 day) may be minimal.  Up to now, differences in the time-delay of the tidal bulge as a function of libration period inferred from LLR have been interpreted as a frequency dependence of the tidal $Q$ factor \cite[][]{williams01,williams14,williams15}.  But taking into account viscoelastic inner core deformation may alter this $Q$ versus frequency relationship.  In turn, this has important implications to our understanding of the rheology of the lunar mantle and LVZ.

Lastly, as Figure \ref{fig:impnsd} shows, if the inner core viscosity is smaller than $10^{14}$ Pa s, the magnitude of the tilt angle of the instantaneous inner core figure with respect to the mantle is significantly decreased.  The periodic, 27.212 day, degree 2 order 1 gravity signal associated with the inner core is directly proportional to this tilt angle \cite[][]{williams07}.  This signal has not yet been detected conclusively \cite[][]{williams15b}, which may be for a variety of reasons, including that the inner core is too small or the density contrast at the ICB too weak.  But it may also be because the tilt angle of the inner core is too small.  The latter may be a consequence of viscoelastic relaxation of the inner core, realigning its shape with the mantle, as we have shown in our study.


\appendix 
\section{Elements of the linear system of equations}

\subsection{The full system of equations}

Defining the quantities

\begin{align}
\lambda_1 & = 1 + \omega + \Phi_2 \, ,\\
\lambda_2 & = 1 + \omega - \alpha_2 \, , \\
 \end{align}
the right-hand side vector $\textbf{y}$ and matrix $\boldsymbol{\mathsf{M}}$ of the linear system of equations 

\begin{equation}
\boldsymbol{\mathsf{M}} \cdot \textbf{x} = \textbf{y} \, , \label{eqa:5by5}
\end{equation}
are

\begin{equation}
\textbf{y} = \begin{bmatrix} - \beta \Phi_1 + (1 + \omega ) S_{11} \Phi_1 + 3 \frac{n^2}{\Omega_o^2} {\cal M} \Phi_1 Re[S_{11}]+ i Im[S_{11}] \Phi_1^t \\
\omega S_{21} \Phi_1\\
 - \beta_s \alpha_3 \Phi_1 + \lambda_2 S_{31} \Phi_1 - e_s \alpha_3 \alpha_g S_{41} \Phi_1 + 3 \frac{n^2}{\Omega_o^2} {\cal M} \alpha_3 \Phi_1 Re[S_{31}] +i Im[S_{31}] \alpha_3 \Phi_1^t  \\
0 \\
0
\end{bmatrix} \, ,
\end{equation}

\begin{equation}
\boldsymbol{\mathsf{M}} = \begin{bmatrix} M_{11} & M_{12} & M_{13} & M_{14} & M_{15} \\
M_{21} & M_{22} & M_{23} & M_{24} & M_{25} \\
M_{31} & M_{32} & M_{33} & M_{34} & M_{35} \\
0 & 0 & 1 & \omega & 0 \\
1 & 0 & 0 & 0 & (1 + \omega)
\end{bmatrix} \, ,
\end{equation}
where the elements $M_{11}$ through to $M_{35}$ are

\begin{subequations}

\begin{equation}
M_{11} = \omega - e + \lambda_1 S_{11} + \frac{\bar{A}_s}{\bar{A}} \alpha_3 S_{31} \Phi_2 \, ,
\end{equation}

\begin{equation}
M_{12} = (1 + \omega)  \frac{\bar{A}_f}{\bar{A}}  + \lambda_1 S_{12} +  \frac{\bar{A}_s}{\bar{A}} \alpha_3  S_{32} \Phi_2 \, ,
\end{equation}

\begin{equation}
M_{13} = (1 + \omega) \frac{\bar{A}_s}{\bar{A}} + \lambda_1 S_{13} + \frac{\bar{A}_s}{\bar{A}} \alpha_3  \bigg( S_{33} \Phi_2 - i Im[S_{31}] \Phi_3^t \bigg)\, ,
\end{equation}

\begin{equation}
M_{14} =  \frac{\bar{A}_s}{\bar{A}} \alpha_3 \bigg( (1 + \omega) e_s + \Big(\beta_s + \alpha_2 S_{33} - 3 \frac{n^2}{\Omega_o^2} {\cal M} Re[S_{31}] \Big) \Phi_2 \bigg) + \alpha_2 \lambda_1 S_{13}  \, ,
\end{equation}

\begin{equation}
M_{15} = \beta \Phi_2 - (1 + \omega)  S_{11} \Phi_2 - 3 \frac{n^2}{\Omega_o^2} {\cal M} \Phi_2 Re[S_{11}] - i Im[S_{11}] \Phi_2^t \, ,
\end{equation}

\begin{equation}
M_{21} = \omega (1 + S_{21}) \, ,
\end{equation}

\begin{equation}
M_{22} = 1 + \omega (1 + S_{22}) + e_f + K_{cmb} +  \frac{\bar{A}_s}{\bar{A}_f} K_{icb} \, ,
\end{equation}

\begin{equation}
M_{23} = \omega S_{23} - \frac{\bar{A}_s}{\bar{A}_f} K_{icb} \, ,
\end{equation}

\begin{equation}
M_{24} = -\omega e_s \alpha_1 \frac{\bar{A}_s}{\bar{A}_f}  + \omega \alpha_2 S_{23}  \, ,
\end{equation}

\begin{equation}
M_{25} = - \omega S_{21}  \Phi_2 \, ,
\end{equation}

\begin{equation}
M_{31} = \omega - \alpha_3 e_s + \big( \lambda_2 + \alpha_3 \Phi_2 \big) S_{31} - e_s \alpha_3 \alpha_g S_{41} \, ,
\end{equation}

\begin{equation}
M_{32} = \alpha_1 e_s + \big( \lambda_2 + \alpha_3 \Phi_2 \big) S_{32} - e_s \alpha_3 \alpha_g S_{42} - K_{icb} \, ,
\end{equation}

\begin{equation}
M_{33} = 1 + \omega + \big( \lambda_2 + \alpha_3 \Phi_2 \big) S_{33} - e_s \alpha_3 \alpha_g S_{43} + K_{icb} -i Im[S_{33}] \alpha_3 \Phi_3^t \, ,
\end{equation}

\begin{equation}
M_{34} = \lambda_2 e_s  + \beta_s \alpha_3 \Phi_2 + \big(  \alpha_2 \lambda_2 + \alpha_2 \alpha_3 \Phi_2 \big) S_{33} - e_s \alpha_3 \alpha_g \alpha_2  S_{43}  - 3 \frac{n^2}{\Omega_o^2} {\cal M} \alpha_3 \Phi_2 Re[S_{31}]\, ,
\end{equation}

\begin{equation}
M_{35} =  \beta_s \alpha_3 \Phi_2 -\lambda_2 S_{31} \Phi_2 +   e_s \alpha_3 \alpha_g S_{41} \Phi_2 - 3 \frac{n^2}{\Omega_o^2} {\cal M} \alpha_3 \Phi_2 Re[S_{31}] -i Im[S_{31}] \alpha_3 \Phi_2^t \, .
\end{equation}
\end{subequations}

\subsection{The system of equations for no inner core}

The linear system of equations in the absence of an inner core is given by 

\begin{equation}
\boldsymbol{\mathsf{M}}' \cdot \textbf{x}' = \textbf{y}' \, ,\label{eqa:3by3}
\end{equation}
where 

\begin{equation}
\textbf{y}' = \begin{bmatrix} - \beta \Phi_1 + (1 + \omega ) S_{11} \Phi_1 + 3 \frac{n^2}{\Omega_o^2} {\cal M} \Phi_1 Re[S_{11}] + i Im[S_{11}] \Phi_1^t \\
\omega S_{21} \Phi_1\\
0
\end{bmatrix} \, ,
\end{equation}

\begin{equation}
\boldsymbol{\mathsf{M}}' = \begin{bmatrix} M'_{11} & M'_{12} & M'_{13}  \\
M'_{21} & M'_{22} & M'_{23} \\
1 & 0  & (1 + \omega)
\end{bmatrix} \, ,
\end{equation}
where the elements $M'_{11}$ through to $M'_{23}$ are

\begin{subequations}

\begin{equation}
M'_{11} = \omega - e + \lambda_1 S_{11}  \, ,
\end{equation}

\begin{equation}
M'_{12} = (1 + \omega)  \frac{\bar{A}_f}{\bar{A}}  + \lambda_1 S_{12}  \, ,
\end{equation}

\begin{equation}
M'_{13} = \beta \Phi_2 - (1 + \omega)  S_{11} \Phi_2 - 3 \frac{n^2}{\Omega_o^2} {\cal M} \Phi_2 Re[S_{11}]- i Im[S_{11}] \Phi_2^t \, ,
\end{equation}

\begin{equation}
M'_{21} = \omega (1 + S_{21}) \, ,
\end{equation}

\begin{equation}
M'_{22} = 1 + \omega (1 + S_{22}) + e_f + K_{cmb}  \, ,
\end{equation}

\begin{equation}
M'_{23} = - \omega S_{21}  \Phi_2 \, .
\end{equation}

\end{subequations}

\acknowledgments
Comments and suggestions by Rose-Marie Baland and an anonymous reviewer helped to improve this paper.  Figures were created using the GMT software \cite[]{gmt}. The source codes, GMT scripts and data files to reproduce all figures are freely accessible at \cite{dumberry_organowski20}.  This work was supported by an NSERC/CRSNG Discovery Grant.


\end{document}